\documentclass[twocolumn,aip,jcp,showpacs,preprint,amsmath,amssymb,showkeys,10pt]{revtex4-1}    

\usepackage[utf8]{inputenc}
\usepackage[TS1,OT1,T1]{fontenc}
\usepackage[english]{babel}
\usepackage{lmodern} 

\usepackage{array}
\usepackage{amsmath}
\usepackage{amssymb}
\usepackage{amsthm}
\usepackage{dsfont}
\RequirePackage{pgfplots}  
\usepackage{epstopdf}
\usepackage{epsfig}
\usepackage{placeins}
\usepackage{setspace}
\usepackage{subfig}
\usepackage{xfrac}

\newcommand\N{\mathbb{N}}

\newcommand\R{\mathbb{R}}

\newcommand\dd{\mbox{d}}

\newcommand\C{\mathcal{C}}
\newcommand\Ct{\widetilde{\mathcal{C}}}
\newcommand\Cc{\mathcal{C}_{\mathrm{cut}}}
\newcommand\Rc{R_{\mathrm{cut}}}
\newcommand\fc{f_{\mathrm{cut}}}

\newcommand\Sn{\mathcal{S}}
\newcommand\ka{\kappa^{-1}}

\newcommand\rhoq{\rho_{\{q_1,\hdots,q_n\} } }
\newcommand\e{\mathrm{e}}
\newcommand\dps{\displaystyle}

\newcommand\soap{\textsc{Soap}}
\newcommand\rmsd{\textsc{Rmsd}}
\newcommand\fracc{\textsc{Frac}}
\newcommand\acd{\textsc{Acd}}
\newcommand\vlj{V_{\mathrm{LJ}}}

\begin{document}

\title{Permutation-invariant distance between atomic configurations}

\author{Gr\'egoire Ferr\'e}
\author{Jean-Bernard Maillet}
\affiliation{CEA, DAM, DIF, F-91297 Arpajon, France}

\author{Gabriel Stoltz}
\affiliation{Universit\'e Paris-Est, CERMICS (ENPC), INRIA, F-77455 Marne-la-Vall\'ee, France}

\date{\today}

\begin{abstract}
We present a permutation-invariant distance between atomic configurations, defined through a functional representation of atomic positions. This distance enables to directly compare different atomic environments with an arbitrary number of particles, without going through a space of reduced dimensionality (\textit{i.e.} fingerprints) as an intermediate step. Moreover, this distance is naturally invariant through permutations of atoms, avoiding the time consuming associated minimization required by other common criteria (like the \emph{Root Mean Square Distance}). Finally, the invariance through global rotations is accounted for by a minimization procedure in the space of rotations solved by Monte Carlo simulated annealing. A formal framework is also introduced, showing that the distance we propose verifies the property of a metric on the space of atomic configurations. Two examples of applications are proposed. The first one consists in evaluating faithfulness of some fingerprints (or descriptors), \textit{i.e.} their capacity to represent the structural information of a configuration. The second application concerns structural analysis, where our distance proves to be efficient in discriminating different local structures and even classifying their degree of similarity.
\end{abstract}

\maketitle


\section{Introduction}

Since one decade, the comparison of two atomic structures has raised a
growing interest
in the fields of biology, physics and chemistry. The term <<atomic structure>>
embodies a wide variety of situations (molecules, crystal, fluids, etc) and refers
in a general sense to a set of atoms. For a molecule, it corresponds to the positions
of its constituting atoms. For a condensed matter system, 
it may be the positions of a given atom and its neighbors.
The methodologies developed to compare two sets of atoms draw their diversity 
from the variety of considered
applications, for example comparing molecules\cite{Karakoc15072006,hopping2004,hopping2010} 
or crystal structures \cite{Karakoc15072006,schutt2014,SadeghiMetric2013,ogano2009}, 
performing the Minima Hopping method
\cite{hopping2004,hopping2010}
and more recently Machine Learning 
approaches used for numerical potentials and forces
\cite{GAPAnderson,Molecular2013,seko2014,GAPTungsten,behler2014ref,handley2014next,Bartok2010GAP}.

In all these fields, efforts have been made
to give a permutation, rotation and translation invariant
measure of similarity (or distance) between atomic configurations. Indeed, it is of paramount
importance to provide a comparison of structures having these invariances, since the studied properties
do not depend on the ordering of atoms and the orientation of axes.  
In biology, rotations are defined along axes passing through
the center of mass of the molecule, whereas for numerical potentials it is
around the atom for which we aim at calculating the energy or forces.
For example, if a particle 
has an energy that depends on all its neighbors, applying a global 
rotation to the system does not change this energy.
In the same way, the local structure of a crystal (such as Face Center Cubic, Cubic 
Center, etc) does not depend on the choice of axes or the indexing of neighbors.

In the field of Machine Learning and crystal recognition, 
several approaches rely on the use
of functions often called \emph{fingerprints} or sometimes
descriptors, which represent the structural information
of an atomic configuration. A comparison
between structures then reduces to the comparison of their associated
fingerprints. 
A possible choice of fingerprints is based on 
the eigenvalues of matrices depending on inter-atomic distances 
between the atoms in the system~\cite{SadeghiMetric2013}. In this framework,
several matrices depending on distances between neighboring atoms can be 
used, such as Weyl matrices  \cite{Bartok2013repr}, Coulomb matrices 
\cite{schutt2014,Rupp2012} or Kohn-Sham Hamiltonian matrices, 
Hessian matrices or overlap matrices \cite{SadeghiMetric2013}.  
Other representations can be
used, such as the symmetry functions of Behler and Parrinello 
\cite{behler2007,behler2011atom,behler2014ref}, bond order parameters\cite{SNR83}, power spectrum, bi-spectrum
and 4D-bi-spectrum \cite{Bartok2013repr,bartokthesis}, or even bit-strings
\cite{flower1998}. Nevertheless, beyond the fact that these representations are 
often arbitrary, they are generally intrinsically incomplete. Indeed, if a system is constituted
of $n$ atoms, it has $3n$ degrees of freedom. All the methods relying on eigenvalues of matrices
\cite{SadeghiMetric2013,schutt2014,Rupp2012} compare at most $n$ eigenvalues, and cannot entirely
represent the system\cite{Bartok2013repr}. Therefore, it is in general delicate
to show that calculating the distance between fingerprints corresponds
 to a genuine distance in a mathematical sense.

Another method used to define a distance between environments is  
the \rmsd{} (\textit{Root Mean Square Distance}) 
\cite{Kabsch1976,SadeghiMetric2013}. This distance is simply the
square root of the sum of square distances between the atoms in each environment.
It is indeed a distance from a mathematical viewpoint, but 
suffers from two main drawbacks. First, the two 
configurations must have the same number of particles; this 
restriction is acceptable when comparing molecules but it is 
unacceptably restrictive in the context of structural analysis
of condensed matter systems. 
Moreover, this distance is not permutation invariant, which 
means that all permutations 
should be tried in order to compare two configurations, 
and for each permutation the optimal rotation should be calculated. 
This is feasible with some advanced Monte Carlo approach 
\cite{SadeghiMetric2013}, but the calculations cannot be carried out 
for large systems. 

The limitations induced by these methodologies call for new efforts 
to define a distance that can be applied to 
large systems with an arbitrary number of atoms, for subsequent use in
Machine Leaning methods \cite{Bartok2013repr,behler2014ref}. In this context,
a direct measure of similarity has been introduced by Bartok
\cite{Bartok2013repr}, the SOAP 
(\textit{Smooth Overlap of Atomic Positions}). 
It does not depend on any fingerprint and does not
require to explore the set of permutations. 
In this article, we generalize the notion of representing a 
configuration by a probability density, and define new permutation-free
distances between configurations by appropriately generalizing \soap{}
and \rmsd. 
We introduce in Section~\ref{sec:frame}
a formalism from which we derive a distance between configurations.
The first task is to give a mathematical definition of what we will call
 \emph{environment} and \emph{configuration} in Section~\ref{sec:conf}.
We discuss in Section~\ref{sec:frac} the Functional Representation of an Atomic
Configuration (\fracc), from which we derive a distance on the space of
configurations in Section~\ref{sec:dist}, first for single chemical element systems, and then for
multi-elements ones. Several extensions of the methods are also proposed
in Section~\ref{sec:generalization}.
We next provide  in Section~\ref{sec:appli} some examples of applications.
The first one is the evaluation of the faithfulness of fingerprints to represent the structure
of a chemical atomic environment. The second application aims at proving the efficiency 
of this distance in the context of structural analysis.

\section{Framework}
\label{sec:frame}
\subsection{Atomic Environment and Configuration}
\label{sec:conf}

We first give a mathematical framework to what is usually called 
an atomic configuration or environment, in order to define a distance on such a space. 
Indeed, a configuration is usually
considered as a simple set of vectors $(q_i)_{i=1}^n$, $q_i\in\R^3$, as in the \textsc{Rmsd} 
framework for example \cite{Kabsch1976}. Nevertheless, this representation is
not adapted when dealing with condensed matter systems, since the number
of neighbors may vary. 
In the following, $\Sn$ is
the space of permutations and $\mathcal{R}$ the space of 
rotations (we use the same notation for rotations of $\R^3$ and rotations of a 
rigid body), \textit{i.e.} matrices $R$ such that $R^T R=RR^T=I$ and $\mathrm{det}\, R=1$. 
For now, we restrict ourselves to systems constituted of a \emph{single} 
chemical element, an extension to multi-elements systems being provided in 
Section~\ref{sec:multi}. 

First, an environment constituted of $n$ atoms is
a set of $n$ vectors of $\R^3$. We therefore define
\[
\C^n=\left\{ \left. (q_i)_{i=1}^n\, \right| \,  \forall \, i\, , \ q_i\in\mathbb{R}^{3} \right\}.
\]
However, we want to consider environments with an arbitrary number
of atoms. One possible appropriate definition of an atomic 
environment is then
\begin{equation}
\label{eq:C}
\C= \underset{n \geq 1}{\bigcup} \C^n.
\end{equation}
The positions are defined with respect to the center of frame which
depends on the application at hand.
This allows to get rid of spurious translation invariances. 

In order to describe an environment with particles in a radius 
$\Rc$ around the origin, we restrict the set of admissible environments to environments having
all their elements within a distance $\Rc$ of the origin:
\begin{equation}
\label{eq:Ccut}
\Cc= \underset{n \geq 1}{\bigcup} \Cc^n,
\end{equation}
with
\[
\Cc^n=\left\{(q_i)_{i=1}^n\, \mid \,  \forall \, i\, , \ \ q_i\in\mathbb{R}^{3}\, , \
 \| q_i \| \leq \Rc  \right\}.
\]
We will see that only small adaptions are 
required to use $\Cc$ rather than $\C$. We therefore work in $\C$ with no restriction, 
and make precise the adaptions required to work with $\Cc$ when necessary.

\begin{figure}[h]
  \centering
      \includegraphics[width=0.9\linewidth]{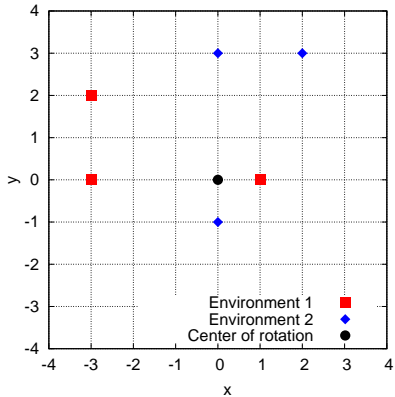}
      \caption{Example of two environments $C_1$ and $C_2$ representing the same configuration $\widetilde{C}$ in two dimensions.}
      \label{fig:env}
\end{figure}

Several atomic properties like the potential energy in classical simulations,
or the local structure, are defined with respect to the environment 
(\textit{i.e.} the set of surrounding particles) of a given atom.
As these properties do not depend on global translations, rotations
or permutations (or equivalently on an arbitrary choice of axes and 
ordering of particles), an appropriate definition of an environment
should retain these invariances.
For an environment $C=(q_i)_{i=1}^{n}\in\mathcal{C}$, a permutation 
$\sigma\in \Sn$, a rotation $R\in\mathcal{R}$, 
we define $\sigma C =(q_{\sigma(i)})_{i=1}^{n}$ and $R C =(Rq_i)_{i=1}^{n}$.
Two environments $C_1$ and $C_2$ are considered equivalent if one is a rotation and/or
a permutation of the other. This suggests the 
following equivalence relationship: for 
$C_1=(q_i)_{i=1}^{n_1}\in\mathcal{C}$, $C_2=(q'_i)_{i=1}^{n_2}\in\mathcal{C}$, 
\[
C_1\sim C_2\Leftrightarrow 
\left| \begin{aligned} 
  & n_1 = n_2\ \mathrm{and} \ \exists \, \sigma \in \mathcal{S}, \, \exists \, R\in \mathcal{R} \\ 
  & \mbox{ such that } C_1=\sigma R C_2.
\end{aligned} \right.
\]
To define what we call a configuration in
the following, we gather in classes all the environments that are equivalent.
As a result, the space of configurations reads:
\begin{equation}
\label{eq:Ct}
\Ct=\C_{/ \sim}.
\end{equation}
This means that each element $\widetilde{C}\in \Ct$ is the ensemble 
of all possible rotations and permutations of an environment. 
From now on, we will call \emph{environment} an element of $\C$ and 
\emph{configuration} an element of $\Ct$.
Once again, two environments $C_1$ and $C_2$ that differ by a rotation 
or a permutation belong to the same class (configuration) $\widetilde{C}$,
and are therefore understood as identical in the sense of configurations.
In practice, a configuration $\widetilde{C}\in\Ct$ can be represented
by any of its elements $C\in \widetilde{C}$
(an example of two environments representing the same configuration is displayed
in Figure~\ref{fig:env}). This definition is motivated
by the fact that in many applications, for example calculating potentials or
comparing molecules, only the \emph{structure} of the environment matters, 
and not its orientation
with respect to arbitrary axes or the ordering of its elements.

We now define a functional representation of an environment. This representation
turns out to be permutation invariant, and is a fundamental tool to develop
in Section~\ref{sec:dist} a distance on $\Ct$, or equivalently a permutation
and rotation invariant distance on $\C$.

\subsection{Functional Representation of Atomic Configuration (\fracc)}
\label{sec:frac}
\begin{figure}
  \centering
      \subfloat[][Atomic environment with central atom (blue) and neighbors (red)]{\includegraphics[width=0.45\linewidth]{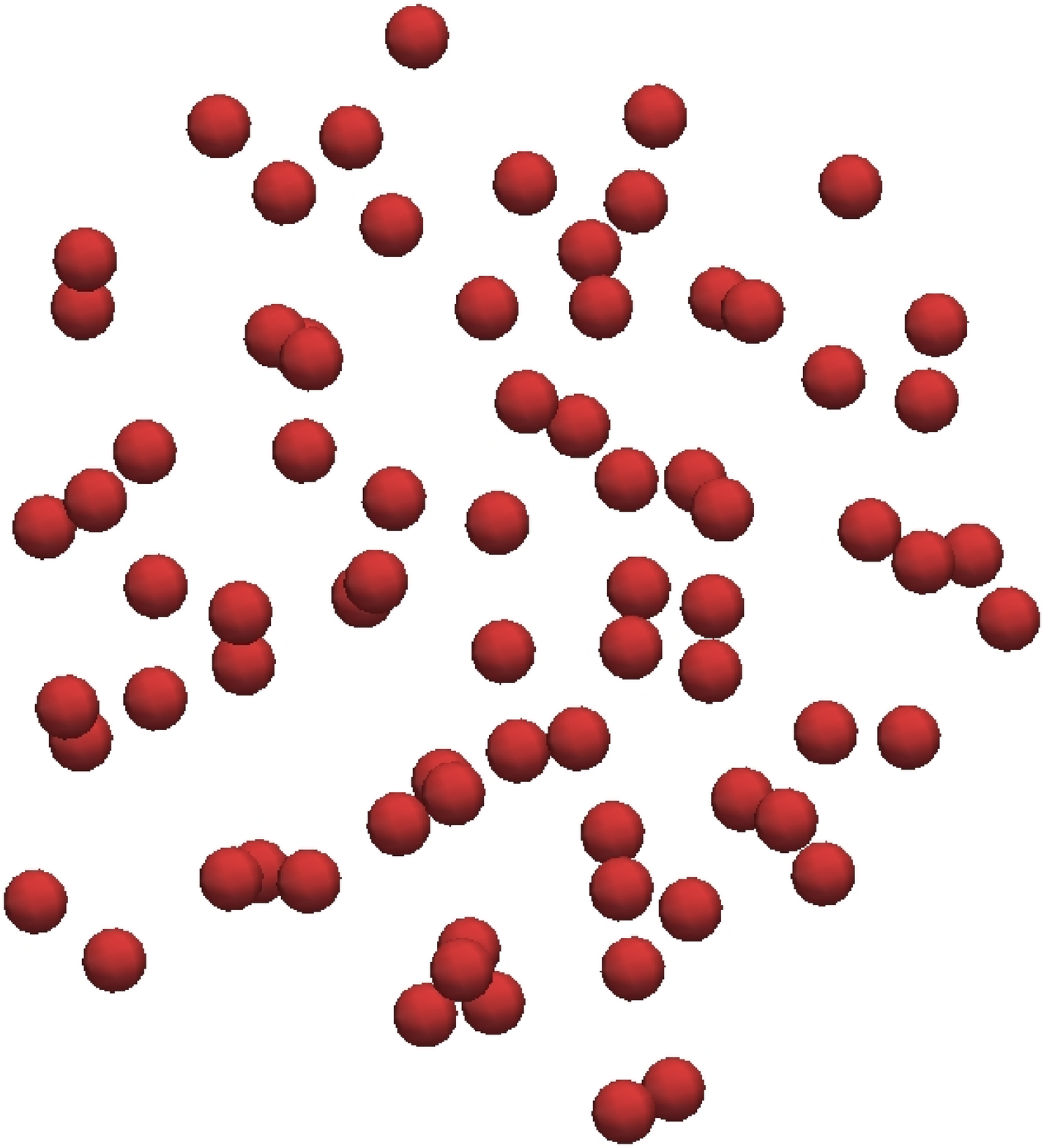}}
      \hspace{5mm}
      \subfloat[][Associated functional representation]{\includegraphics[width=0.45\linewidth]{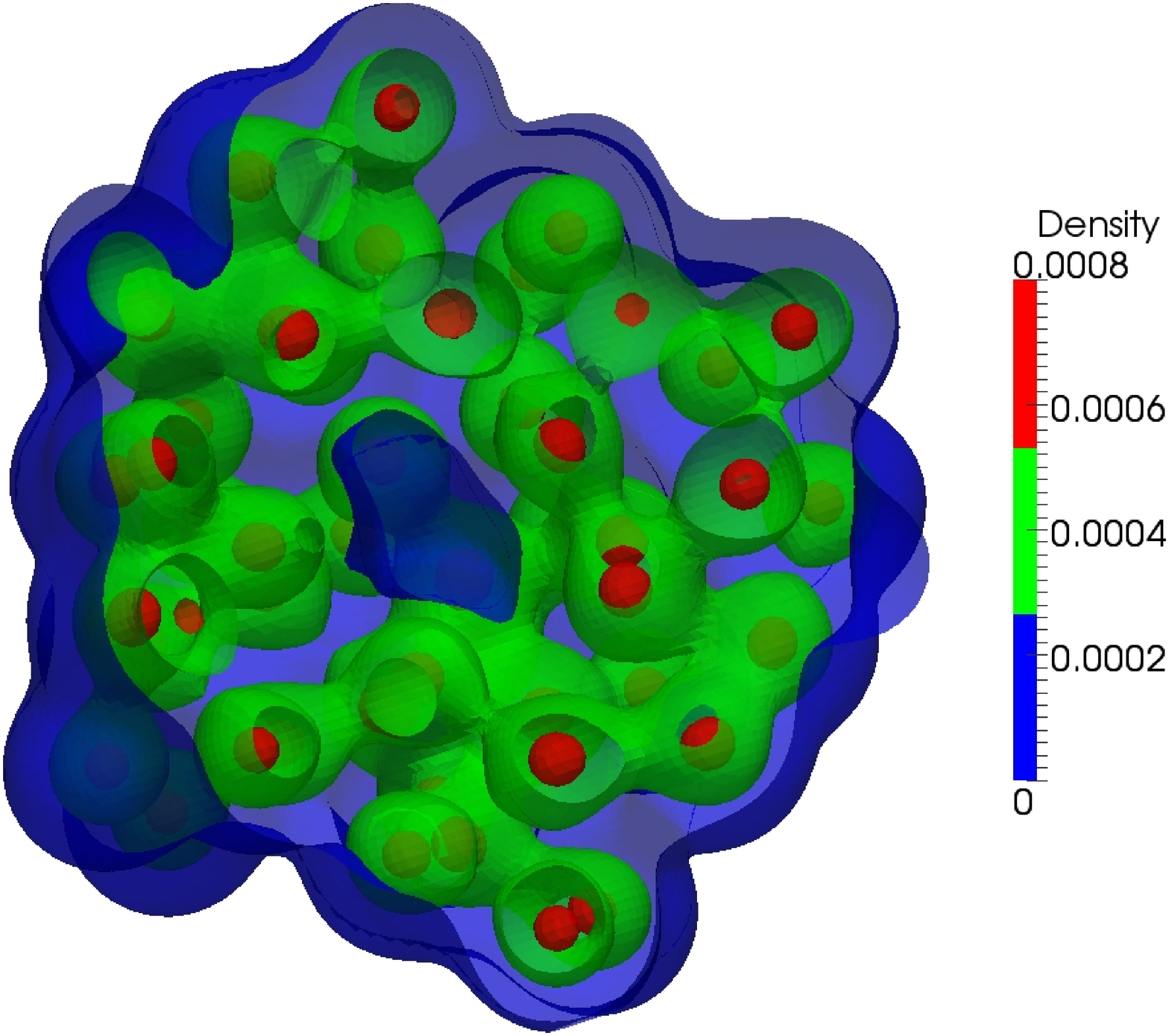}}
      \hspace{5mm}
      \caption{Example of atomic environment for a fluid with associated \fracc}
      \label{fig:density}
\end{figure}

In this part we still consider systems with a single chemical element. Our goal
is to represent a set of particles by a smooth probability density that can
be manipulated more conveniently than a set of vectors. 
The idea is that a set of $n$ particles at positions $(q_i)_{i=1}^n$ can be 
exactly represented 
by a set of Dirac functions $\delta_{q_i}$ at their positions:
\begin{equation}
\label{eq:dirac}
\rhoq=\frac{1}{n}\sum_{i=1}^{n}\delta_{q_i}.
\end{equation}
This expression is convenient since it is invariant with respect to
permutation of atoms: we exploit this property in Section~\ref{sec:dist}. 
But this representation is unusable in practice,
because the spectral expansion of a Dirac delta function is too slowly convergent. Approximating 
the Dirac delta function requires basis sets of very large dimension. 
It is more convenient to smooth out~(\ref{eq:dirac}) and 
represent an environment by a density $\rho_{\sigma}$, where $\sigma$ is a
smoothness parameter, 
as in Refs.~\onlinecite{Bartok2013repr,bartokthesis}. This density $\rho_{\sigma}$ can be
interpreted as a density 
of mass, or as a probability density 
of presence, the particles' positions being considered as realizations of a 
random variable in $\R^3$. 
As a result, we need to define appropriate functions $\varphi$ to represent 
the presence of one atom at position $0$ (other locations are obtained by translations
of such functions). 
These functions should verify the following properties:
\begin{itemize}
\item regularity, e.g. $\varphi$ is continuous by parts;
\item $\varphi$ is bounded with unit mass, \textit{i.e.} $0\leq \varphi \leq 1$
and $\int_{\R^3} \varphi =1$;
\item $\varphi$ is positive and reaches its maximal value at $q=0$;
\item $\varphi(q) \xrightarrow[q \to +\infty]{} 0$.
\end{itemize}
Note that we could restrict ourselves to more regular functions, for example 
infinitely differentiable ones. Several choices can be made, as made precise
in section~\ref{sec:generalization}, but for the sake of clarity we
consider in the sequel the Gaussian case. The choice \cite{Bartok2013repr}
\begin{equation}
\label{eq:gaussian}
\varphi^{\sigma}(q)= (2\pi \sigma^2)^{- \frac{3}{2}} \exp \left( - \frac { \| q \|^2 } 
{2\sigma^2} \right)
\end{equation}
verifies the required properties to represent
the presence of an atom.
A configuration $C=(q_i)_{i=1}^{n}
\in\mathcal{C}$ can then be represented by the density
\begin{equation}
\label{eq:density}
\rho_{\sigma}(q)=\frac{1}{n}\sum_{i=1}^{n}\varphi_{q_i}^{\sigma}(q)
=\frac{1}{n}\sum_{i=1}^{n}\varphi^{\sigma}(q-q_i).
\end{equation}
As an example, the functional representation
of an atomic environment of a fluid with such a function
is shown  in Figure~\ref{fig:density}. The asymptotic behavior
of this density shows that the functional representation~(\ref{eq:density}) is a smoothed 
representation of atomic positions. Indeed, the Gaussian function tends to a 
Dirac delta function when $\sigma$ tends to $0$. As a result
\[
\rho_{\sigma}=\frac{1}{n}\sum_{i=1}^{n}\varphi_{q_i}^{\sigma}\
\xrightarrow[\sigma \to 0]{}
\rhoq=\frac{1}{n}\sum_{i=1}^{n}\delta_{q_i},
\]
in the sense of distributions. More details about this functional 
representation are provided 
in section~\ref{sec:generalization}.

In some applications, it is convenient to consider only atoms within a 
cut-off radius of the origin. We introduce to this end a decreasing cut-off 
function $\fc:\R \mapsto \R$ such that
\[
\fc(0)=1\, , \quad \fc(r) =0 \ \mathrm{for} \ r \geq \Rc.
\]
For instance, Parrinello and Behler \cite{behler2007,behler2011atom,behler2014ref} 
choose
\begin{equation}
\label{eq:fcut}
\fc(r) = \left\lbrace
\renewcommand{\arraystretch}{1.5}
\begin{array}{*2{>{\displaystyle}l}}
\frac{1}{2} \left[ \cos \left( \frac{\pi r}{\Rc} \right) 
+1 \right]\, , & r \leq \Rc, \\
0\, ,& r > \Rc.
\end{array} \right.
\end{equation}
Of course other options may be considered, such as $\fc(r)=\mathds{1}_{\{r < \Rc  \} }(r) $. 
In this framework, an environment is now represented by its smoothed weighted density:
\begin{equation}
\label{eq:densitycut}
\rho_{\Rc}(q)=\frac{\dps \sum_{i=1}^{n} 
\varphi_{q_i}(q)\fc(\| q_i \| )}{\dps \sum_{i=1}^{n} \fc(\| q_i \|)}.
\end{equation}
The denominator in~(\ref{eq:densitycut}) is introduced in order for $\rho_{\Rc}$ 
to be a probability density (\textit{i.e.} with unit mass). From a physical 
viewpoint, it shows that we do not attach the same weight to each particle: 
atoms far away from the center have a smaller weight since they should contribute
less to the description of the environment. It represents 
an effective number of particles in the neighborhood of the considered atom. 
All the following results are presented with the representation~(\ref{eq:density})
but can be easily adapted with~(\ref{eq:densitycut}).

Let us emphasize some advantages of the \fracc. First, it is invariant
with respect to permutations of atoms of an environment. Secondly,
it provides the same type of representation (a function on $\R^3$) 
whatever the number of atoms, which allows to compare configurations
with different numbers of atoms. Finally, a distance between configurations
can be naturally defined in this functional framework.

\section{Atomic Configuration Distance (\acd)}

\label{sec:dist}
\subsection{Single Element Systems}
\label{sec:distsingle}
Once a \textsc{Frac} has been defined for systems with one chemical element, 
a natural way to derive a distance between configurations
is to define a distance between their associated densities. A classical choice
is the $L^2$ distance since it relies on a standard scalar product. Indeed, 
defining as in Ref.~\onlinecite{Bartok2013repr} the overlap integral
\begin{equation}
\label{eq:Sint} 
S(\rho_1,\rho_2)=\int_{\R^3} \rho_1 \rho_2, 
\end{equation}
the $L^2$ distance reads
\begin{equation}
\label{eq:l2}
\|\rho_1 - \rho_2\|_{L^2}^2=S(\rho_1,\rho_1) -2S(\rho_1,\rho_2) + S(\rho_2,\rho_2).
\end{equation}
By identification between an environment and its density, we define 
the following distance, for $C_1=(q_i)_{i=1}^{n_1}$, and $C_2=(q'_i)_{i=1}^{n_2}
\in\mathcal{C}$:
\begin{equation}
\renewcommand{\arraystretch}{2.5}
\begin{array}{*2{>{\displaystyle}l}}
d_2\left(C_1,C_2\right)^2& = \frac{1}{n_1^2}\, \sum_{i,j=1}^{n_1}
  \int_{\R^3} \varphi_{q_i}(q)\varphi_{q_j}(q) \dd q \\
&-\, \frac{2}{n_1 n_2}\, \sum_{i=1}^{n_1} 
\sum_{j=1}^{n_2}  \int_{\R^3} \varphi_{q_i}(q)\varphi_{q'_j}(q) \dd q
 \\ & +\, \frac{1}{n_2^2}\, \sum_{i,j=1}^{n_2} 
\int_{\R^3} \varphi_{q'_i}(q)\varphi_{q'_j}(q)\dd q.
\end{array}
\label{eq:funcACD}
\end{equation}
In order to construct a rotation invariant distance, we
take the infimum over rotations of one representing environment,
as done for the \rmsd{} \cite{coutsias2004,Theobald2005}. 
For two configurations $\widetilde{C}_1$, $\widetilde{C}_2\in\Ct$,
we choose two representing environments $C_1\in \widetilde{C}_1$ and 
$C_2\in \widetilde{C}_2$ belonging to each of these configurations and
define:
\begin{equation}
\label{eq:metric}
\renewcommand{\arraystretch}{2}
\begin{array}{*2{>{\displaystyle}l}}
\tilde{d_2}\left(\widetilde{C}_1,\widetilde{C}_2\right) & =
\underset{R\in\mathcal{R}}{\inf}\, d_2\,(C_1,RC_2) \\
& =\underset{R\in\mathcal{R}}{\inf}\, d_2\,(RC_1,C_2),
\end{array}
\end{equation}
which is a distance on $\Ct$ (see Appendix~\ref{appendix:A}). In practice, we 
compare one environment $C_1$ to another $C_2$ with~(\ref{eq:funcACD})
and estimate the rotation minimizing this quantity, as in a \rmsd{}
framework \cite{Kabsch1976}. Nevertheless, there seems to be no simple
expression of this optimal rotation, contrarily to the \rmsd{} case \cite{coutsias2004,Theobald2005}
for which the optimal rotation is obtained as solution of a singular
value decomposition. We describe a procedure to numerically estimate 
the optimal rotation in section~\ref{sec:num}.

The \emph{Smooth Overlap of Atomic Positions} (\soap) introduced 
in Ref.~\onlinecite{Bartok2013repr} also consists in taking a Gaussian function for 
$\varphi$ as in~(\ref{eq:gaussian}) but averaging over rotations 
by decomposing over spherical harmonics. Here, we also choose Gaussians,
but rather consider the infimum over rotations. The idea is drawn from the \rmsd{},
and enables to define a permutation and rotation invariant distance on $\C$, 
or equivalently a 
distance on $\Ct$. Moreover, in the Gaussian case, the overlap integral~(\ref{eq:Sint}) 
has an analytical
expression which alleviates the need for numerical quadrature. 
Indeed, the following formula holds (see Appendix~\ref{appendix:B}):
\begin{equation}
\label{eq:S}
S(\rho_1,\rho_2)= \frac{\ka}{n_1 n_2} \sum_{i=1}^{n_1} 
\sum_{j=1}^{n_2} \exp\left(- \frac{(q_i - q'_j)^2}{4\sigma^2} \right),
\end{equation}
with $\kappa=8 (\pi \sigma^2)^{ \frac{3}{2} }$. The permutation invariant measure of 
distance based on~(\ref{eq:funcACD}) then simplifies as
\begin{equation}
\label{eq:ERMSD}
\renewcommand{\arraystretch}{3}
\begin{array}{*2{>{\displaystyle}l}}
d_2\left(C_1,C_2\right)^2&=  \frac{\ka}{n_1^2}\, \sum_{i,j=1}^{n_1} 
 \exp\left(- \frac{(q_i - q_j)^2}{4\sigma^2} \right) \\
&- \,\frac{2\ka}{n_1 n_2}\, \sum_{i=1}^{n_1} \sum_{j=1}^{n_2}   
\exp\left(- \frac{(q_i - q_j')^2}{4\sigma^2} \right)  \\ 
&+\, \frac{\ka}{n_2^2}\,  \sum_{i,j=1}^{n_2}   \exp\left(- 
\frac{(q_i' - q_j')^2}{4\sigma^2} \right).
\end{array}
\end{equation}
%
%
A permutation and rotation invariant distance is finally obtained with~(\ref{eq:metric}).
This is the formula we use in the applications presented in this work, with the
cut-off function as given in Section~\ref{sec:frac}. Moreover, asymptotic
results when $\sigma\to0$ and $\sigma \to +\infty$ are studied in Appendix~\ref{sec:asympt}.

Finally, as noticed in Ref.~\onlinecite{Bartok2013repr}, the quantity
\[
S(C_1,C_2)=\frac{1}{n_1 n_2} \sum_{i=1}^{n_1} \sum_{j=1}^{n_2} 
\exp\left(- \frac{(q_i - q'_j)^2}{4\sigma^2} \right)
\]
is a \emph{permutation invariant measure of similarity} between atomic
environments. Setting 
\[
\tilde{S}(C_1,C_2)=\frac{ \underset{R\in \mathcal{R}}{\sup} S(C_1,R C_2)}   
{\sqrt{ S(C_1, C_1) S(C_2, C_2)}  },
\]
we define a kernel on the space of configurations $\Ct$ 
that can be directly used for Machine Learning approaches, 
as \soap{}\cite{Bartok2013repr}. The supremum ensures that the optimal
rotation maximizes the similarity of the structures.
\subsection{Optimal Rotation}
\label{sec:num}
From a practical viewpoint,
we search the minimum in~(\ref{eq:metric}) by parameterizing 
a rotation $R$ in terms of the Euler angles\cite{rose1995elementary,varshalovich1988quantum} 
$(\alpha,\beta,\gamma)$ , but other
choices could be made, such as quaternions \cite{coutsias2004,Theobald2005}. The three 
angles correspond to three rotations around different axes. We use rotations 
of angle $\nu$ over axes $x$ and $z$, defined by the following rotation matrices:
\[
R_x(\nu)=
\begin{pmatrix}
1 & 0 & 0 \\
0 & \cos(\nu) & -\sin(\nu) \\
0 & \sin(\nu) & \cos(\nu)
\end{pmatrix} \, , 
\] 
\[
R_z(\nu)=
\begin{pmatrix}
\cos(\nu) & -\sin(\nu) & 0 \\
\sin(\nu) & \cos(\nu) & 0 \\
0 & 0 & 1
\end{pmatrix}.
\]
A rigid body rotation \cite{rose1995elementary,varshalovich1988quantum}
is then defined by $R_{\alpha,\beta,\gamma}=R_z(\gamma) 
R_x(\beta) R_z(\alpha)$ with $(\alpha,\beta,\gamma)\in [0,2\pi]\times [0,\pi] 
\times [0,2\pi]$. Therefore, the optimization is performed in this 3-dimensional
compact space. We use a Monte Carlo procedure
(simulated annealing \cite{SA1987,SA2005}, see Appendix~\ref{sec:siman}) to search for the minimum.

\begin{figure}
  \centering
      \subfloat[][Environment $C_1$]{\includegraphics[width=0.45\linewidth]{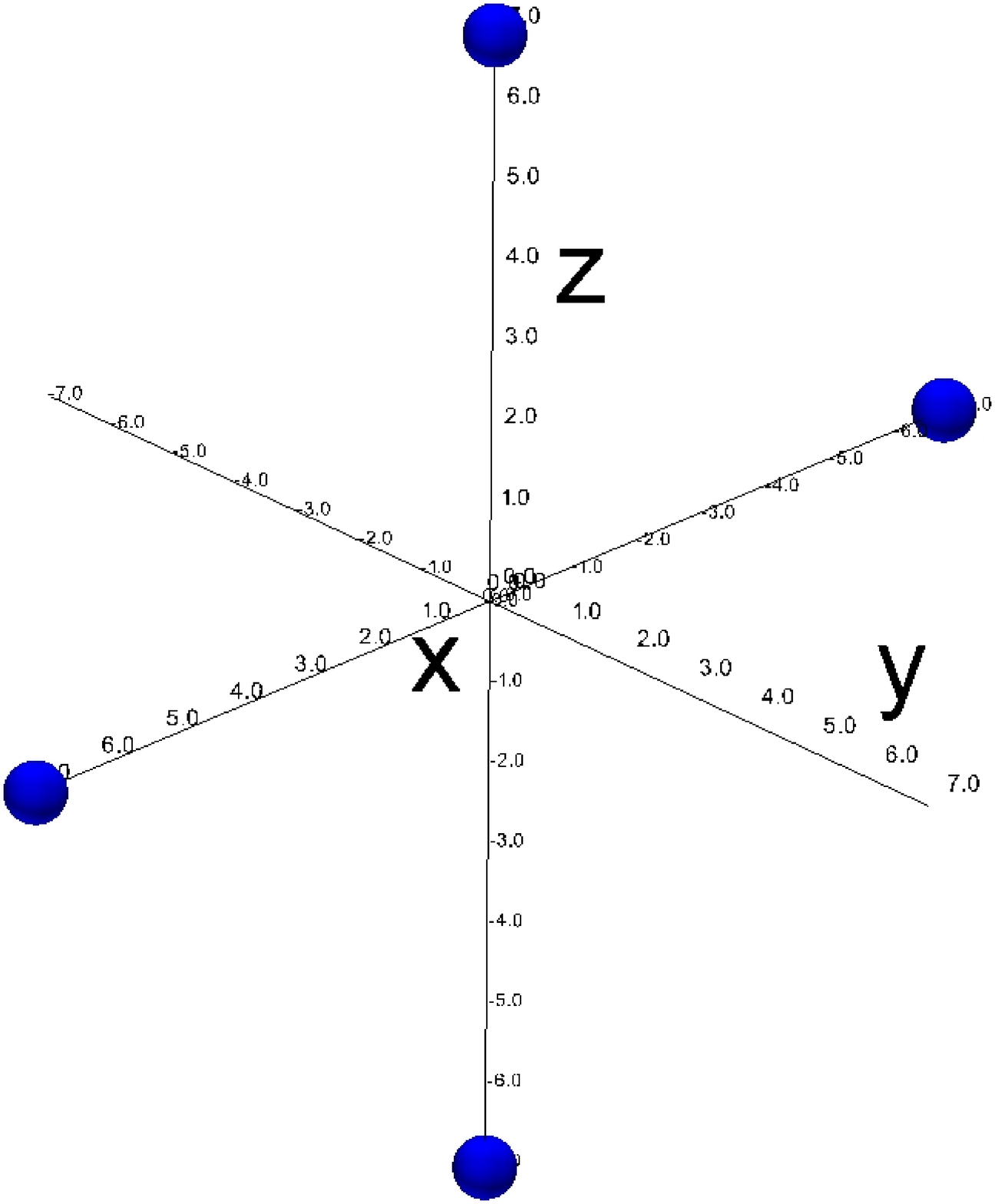}}
      \hspace{5mm}
      \subfloat[][Environment $C_2$]{\includegraphics[width=0.45\linewidth]{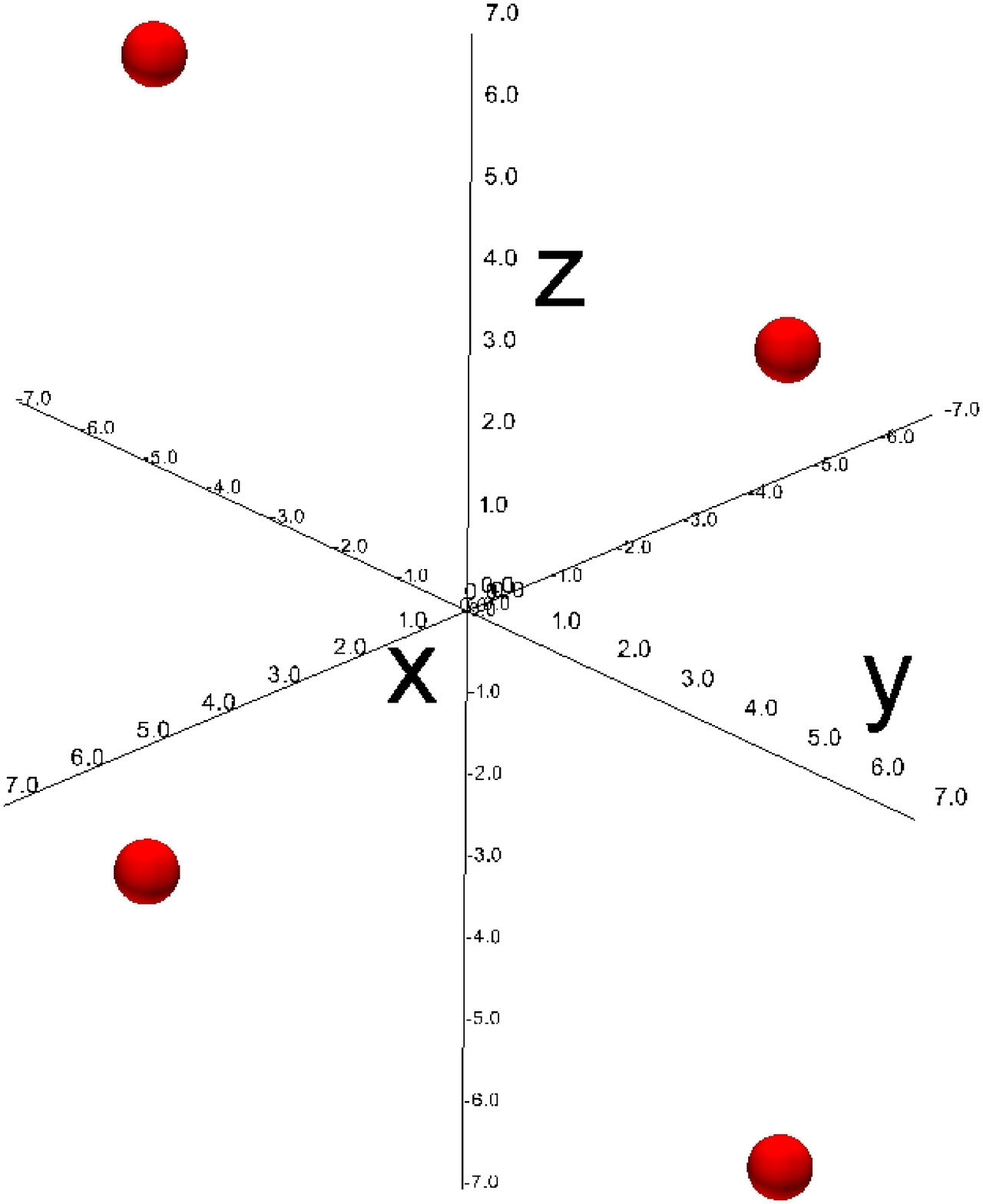}}
      \hspace{5mm}
      \caption{Example of two environments $C_1$ and $C_2$ representing the same configuration $\widetilde{C}$ in three dimensions.}
      \label{fig:env2}
\end{figure}

\begin{figure}[h]
  \centering
      \includegraphics[width=0.9\linewidth]{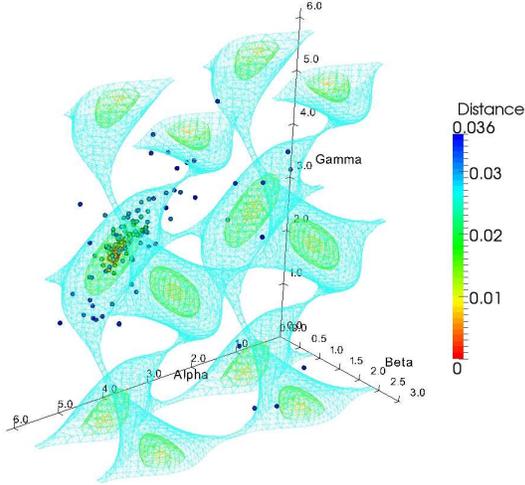}
      \caption{Isovalues of the mapping $J:(\alpha,\beta,
      \gamma)\mapsto d_2(C_1,R_{\alpha,\beta,\gamma}C_2)$ when $C_1$ and $C_2$ belong to the same class. 
      The points represent the iterates of a simulated annealing procedure.}
      \label{fig:isosimple}
\end{figure}

\begin{figure}[h]
  \centering
      \includegraphics[width=0.9\linewidth]{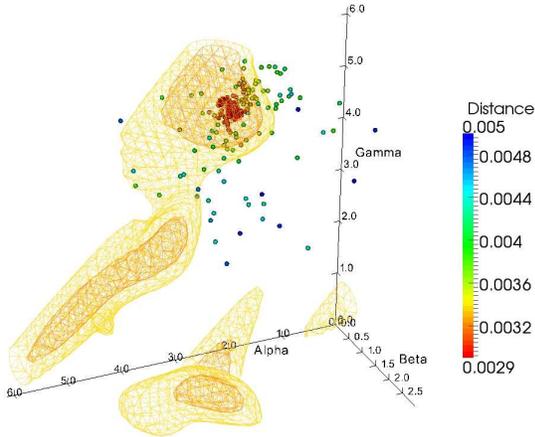}
      \caption{A realistic example, two configurations of a fluid. 
      The wire frame represents level sets of the mapping 
      $J:(\alpha,\beta,
      \gamma)\mapsto d_2(C_1,R_{\alpha,\beta,\gamma}C_2)$
      and the points a simulated annealing procedure.}
      \label{fig:isorea}
\end{figure}

In order to test the method, we fix two environments $C_1$, $C_2$ 
with 4 particles in each one
and $C_1=\sigma RC_2$ for some rotation $R$ and permutation $\sigma$ (see Figure~\ref{fig:env2}),
so that $C_1$ and $C_2$ represent the same configuration. Then we study the
mapping $J:(\alpha,\beta,\gamma)\mapsto d_2(C_1,R_{\alpha,\beta,\gamma}C_2)$.
In Figure~\ref{fig:isosimple},
we display isosurfaces of $J$. 
We see that the minimum of this function
is indeed 0, and that several rotations meet this minimum (given that the system has
several symmetries). Moreover, a realization of simulated annealing 
is plotted, showing that the global minimum is found by a numerical method. 
As a second example, two environments $C_1$, $C_2$ of a Lennard-Jones fluid 
(see Appendix~\ref{sec:SM}) are considered. An example of such an environment is
represented in Figure~\ref{fig:density}~(a).
In Figure~\ref{fig:isorea}, we plot isosurfaces of $J$
with a realization of simulated annealing for these environments.
Here, two local minima are found, but only one is the global minimum.

We studied more specifically the case of the Gaussian \fracc{} with $L^2$ distance,
since the analytical formula~(\ref{eq:ERMSD}) allows to dramatically reduce the computational
cost of its evaluation. In this case, the parameter $\sigma$,
if not essential from a theoretical viewpoint, plays a role in numerical
applications. Indeed, changing $\sigma$ does not change the position of the global
minimum, but affects the gradient of function $J$. A large value of $\sigma$ leads to a shallow
function, whereas a small $\sigma$ leads to a peaked function $J$.
Therefore, the numerical
efficiency of the optimization procedure depends on the value of this
parameter.

\subsection{Generalizations}
\label{sec:generalization}
We considered in the previous section the particular case of Gaussian
functions $\varphi$ for representing atoms.
Let us however emphasize that other choices are possible and in fact many other
 generalizations of \fracc{} can be proposed.

First, several choices are available for representing the presence of 
one atom, for example in dimension~$1$: 
\begin{itemize}
\item $\dps \varphi_1 (q) = (2\pi\sigma^2)^{- \frac{1}{2}  } \exp\left( -\frac{\|q\|^2}{2\sigma^2} \right)$
\item $\dps \varphi_2 (q) = \frac{c}{\sigma} \exp\left(-\frac{1}{\sigma^2-\|q\|^2} \right)$
\item $\dps \varphi_3 (q) = \frac{1}{2\sigma} \mathds{1}_{ \{\| q \| \leq \sigma \} }(q)$
\item $\dps \varphi_4 (q) = \left\{ \begin{array}{*2{>{\displaystyle}l}}
      q + \sigma, & \mbox{for } q \in (- \sigma, 0), \\
    -q + \sigma, & \mbox{for } q \in (0, \sigma), \\
    0, & \mathrm{otherwise}.
       \end{array} \right. $
\end{itemize}

The idea always is to put weight in the vicinity of the central atom, with more or less smoothness and accuracy.
More formally, this reduces to using an approximation of a Dirac 
function in the sense of distributions. 
The indicator function in dimension~3,
$\varphi^{\sigma}(q)=(\frac{4}{3}\pi\sigma^3)^{-1}\mathds{1}_{ \{\| q \| 
< \sigma\} }(q)$, represents the atom as a volume in space. Other functions
represent a mass density vanishing more or less smoothly. Another 
interpretation is probabilistic. We can consider that the positions of neighbors
are realizations of a random variable drawn according to a density $\chi$. In this case, 
the density~(\ref{eq:density}) with the choice $\varphi^{\sigma}(q)=(\frac{4}{3}\pi\sigma^3)^{-1}
\mathds{1}_{ \{\| q \| < \sigma \} }(q)$ corresponds to a histogram 
in $\R^3$ approximating
this measure $\chi$. The Gaussian
choice corresponds to the Gauss kernel approximation of $\chi$, which is a particular
case of non-parametric estimation of a probability density 
\cite{silverman1986density,hardle2004nonparametric}, and can be interpreted as 
a smoothed histogram. 

From a formal point of view, the measure $\rho_{\sigma}$ defined in~(\ref{eq:density})
corresponds to the convolution of the measure $\rhoq$ defined in~(\ref{eq:dirac}) with
the shape function $\varphi^{\sigma}$, which reads:
\[
\renewcommand{\arraystretch}{2.5}
\begin{array}{*2{>{\displaystyle}l}}
\rho_{\sigma}(q)& = \rhoq \star \varphi^{\sigma}(q) \\ &  = \int_{\R^3} \rhoq (q') 
\varphi^{\sigma}(q-q')\dd q'.
\end{array}
\] 
As a result, the function $\varphi^{\sigma}$ can be interpreted either as a way to represent
an atom or as a shape function in a regularizing convolution of the measure $\rhoq$.

Another possible extension of the method is to consider different distances
between densities, such as $L^p(w)$ distances where $p\in \N^*$ and $w$
is typically a rotation invariant non-negative weight function with
unit mass. In this case,
\begin{equation}
\|\rho_1 - \rho_2\|_{L^p(w)} = \left( \int_{\R^3} 
| \rho_1 -\rho_2 |^p w \right)^{\frac{1}{p}}.
\end{equation}
The weight $w$ can be used for example to give a higher
importance to some region of space. In our case, the choice $p=2$
is reasonable not only because it is coherent with the distances generally used
(for example between sets of vectors for the \textsc{Rmsd}), 
but also because it relies on a scalar product that proves to
be convenient for calculations. On the other hand, no analytical expressions are in general available for the 
extensions listed so far. They are, as a result, more computationally
expensive in practice.

It is nonetheless possible to consider another way to generalize \acd, by 
directly starting from~(\ref{eq:S}) and replacing the Gaussian function by 
any shape function $\varphi^{\sigma}$ verifying the properties
outlined in Section~\ref{sec:frac} (typically one presented at the begining of 
this section).
This amounts to considering
\[
S(C_1,C_2)=\frac{1}{n_1 n_2} \sum_{i=1}^{n_1} 
\sum_{j=1}^{n_2}   \varphi^{\sigma} ( \|q_i -q'_j \| ).
\]
The function $\varphi^{\sigma}$ now plays the role of a correlation between
two particles. If the particles are close with respect to the bandwidth $\sigma$, 
their distance is small and $\varphi^{\sigma}$ is typically
close to 1. If they are distant, $\varphi^{\sigma}$ is close to 0.
This opens a wide range of perspectives for constructing distances for which the same conclusions as in 
Section~\ref{sec:dist} can be drawn.

\subsection{Multi-elements Systems}
\label{sec:multi}

We now extend the previous results to systems of $p$ 
chemical elements, which offer a wider range of applications.
In our framework, a natural way to represent an environment of $p$ chemical 
species is to represent it as a product of environments of 
individual species:
\begin{equation}
\label{eq:multiC}
\mathcal{C}^p = \underbrace{ \mathcal{C} \times \hdots \times \mathcal{C}}_{p\, \mathrm{times}}.
\end{equation}
The only issue is to 
correctly describe admissible rotations and permutations of such a system. 
For a multi-element environment $C=(C^1, C^2, \hdots ,C^p) \in \mathcal{C}^p$, 
with $n_i$ elements for each $C^i$, we define a permutation of the system 
$\sigma=(\sigma_1,\sigma_2, \hdots, \sigma_p)\in \mathcal{S}_n= \mathcal{S}_{n_1}
\times \hdots \times \mathcal{S}_{n_p}$, as
\[
\sigma C = \left(\sigma_1 C^1,\sigma_2 C^2, \hdots ,\sigma_p C^p \right),
\]
and a rotation $R\in \mathcal{R}$ as
\[
R C = \left( R C^1,R C^2, \hdots ,R C^p\right).
\]
First, this means that permutations are allowed only between elements of the same
chemical element (and not with other species). Secondly, rotations
are defined on the whole system, so they modify each environment in the same way.
It is then straightforward to define the equivalence relationship $\sim$ 
as in Section~\ref{sec:conf}. For $C_1=((q_i^1)_{i=1}^{n_1},\hdots,
(q_i^{p})_{i=1}^{n_p}) \in\mathcal{C}^p$, $C_2=((q_i'^1)_{i=1}^{m_1},
\hdots,(q_i'^{p})_{i=1}^{m_p}) \in\mathcal{C}^p$, 
\[
C_1\sim C_2\Leftrightarrow \left| \begin{aligned}
& \forall\, i\leq p,\, n_i=m_i \mbox{ and }  \exists \, \sigma \in \mathcal{S}_n, \, \exists \, R\in \mathcal{R}\\ 
& \quad \mbox{such that }\, C_1=\sigma R C_2.
\end{aligned}\right.
\]
We next define $\Ct^p=\C^p_{\, /\sim}$. In other words, two environments belong to the same
class if and only if one can be written as a rotation
of the other up to a permutation of atoms in each species.
Now, we can introduce a distance between 
two multi-elements environments as the sum of distances for each single element, that is:
\[
d_2^p(C_1,C_2)^2 = d_2(C_1^1,C_2^1)^2+ \hdots + d_2(C_1^p,C_2^p)^2,
\]
where $d_2$
is defined~(\ref{eq:ERMSD}) in the Gaussian case.
This measure of distance is permutation invariant thanks to the \fracc{} framework. 
Now, to make this distance rotation invariant, we again minimize
over rotations. For two configurations $\widetilde{C}_1$, 
$\widetilde{C}_2\in \widetilde{\C}^p$, represented by two environments
$C_1\in\widetilde{C}_1$ and $C_2\in\widetilde{C}_2$, we set
\begin{equation}
\label{eq:metricCp}
\tilde{d}^p\left(\widetilde{C}_1,\widetilde{C}_2\right)=\underset{R\in\mathcal{R}}{\inf}\, \sqrt{\,  \sum_{i=1}^{p}\, d_2(C_1^i,R C_2^i)^2 }.
\end{equation}
Note that since rotations act in the same way on all single environment
(\textit{i.e.} for each chemical element), the minimum is taken over
the sum of distances with the same rotation for each species.
If one wants to give various importance to the different 
elements, a family of distances may be introduced with a vector of positive weights 
$w\in(\R_{+}) ^p$ with $\sum_{i=1}^pw_i=1$, by setting
\[
\tilde{d}_w^p\left(\widetilde{C}_1,\widetilde{C}_2\right)=\underset{R\in\mathcal{R}}{\inf}\, \sqrt{\, 
 \sum_{i=1}^{p}\, w_i\, d_2(C_1^i,R C_2^i)^2 }.
\]
We can check that $\tilde{d}_w^p$ verifies all the properties 
of a distance on $\Ct^p$ as in the single element case. 
Of course, an extension to configurations in a cut-off radius $\Cc$ is 
straightforward following the discussion of Section \ref{sec:frac}.

\section{Applications}
\label{sec:appli}

\subsection{Faithfulness of fingerprints}
\label{sec:descriptor}

Fingerprints appear in several areas of chemical physics,
for example for representing Potential Energy Surfaces (PES) 
with Machines Learning approaches \cite{Bartok2013repr,bartokthesis,behler2007,
behler2011atom,behler2014ref} or local structure recognition 
\cite{schutt2014,SadeghiMetric2013,Bartok2013repr,bartokthesis}. 
The problem with 
these  methods is that two environments 
that differ by a rotation or a 
permutation are understood as different inputs for the model, and may
lead to different outputs, whereas the energy and the forces remain unchanged. 
Moreover, the  dimension of the input vector has to be constant,
and should not depend on the number of neighbors. Therefore, fingerprints (or descriptors) generally
consist of functions of neighboring atoms, invariant by rotation
and permutation. They are supposed to characterize the \emph{structure} of
an atomic environment. Formally, they can be defined as a mapping
$\Phi : \C \mapsto \R^m$ such that for any permutation $\sigma \in \mathcal{S} $ 
and rotation $R \in \mathcal{R}$, it holds $\Phi(\sigma R C)=\Phi(C)$.
We already mentioned in the introduction some appropriate descriptors for such
applications, especially the ones based on eigenvalues of matrices
\cite{SadeghiMetric2013,Bartok2013repr,schutt2014,Rupp2012}, bond-order
parameters and bi-spectrum \cite{Bartok2013repr,bartokthesis} and 
symmetry functions \cite{behler2007,behler2011atom,behler2014ref}.

As an example, we chose to study the faithfulness of representation
of  two types of symmetry functions
\cite{behler2007,behler2011atom,behler2014ref}, but all the cited descriptors could
be tested. We investigate the relevance of two-body functions $G^2$
and three-body functions $G^3$ (see (\ref{eq:G2}) and (\ref{eq:G4}) below).
The sums run over the neighboring atoms and $\theta_{jk}$ is the angle 
associated with the triplet made of the central particle and its neighbors $ j, k$. Moreover, $\fc$ 
is chosen as in~(\ref{eq:fcut}) and $R_s$, $\eta$, $\lambda$ 
and $\zeta$ are real parameters. \\

\onecolumngrid
\begin{equation}
\label{eq:G2}
G^2=\sum_{i=1}^{n}\e^{-\eta( \| q_i \|-R_s)^2}\fc(\| q_i \|).
\end{equation}
\begin{equation}
\label{eq:G4}
G^3=2^{1-\zeta} \sum_{j,k=1}^{n}  (1+\lambda \cos (\theta_{jk}))^{\zeta}  
 \e^{-\eta \left( \| q_j \|^2  +\| q_k \|^2+ 
  \| q_j - q_k \|^2 \right) } \fc(\| q_j\|)\fc(\| q_k \|)\fc(\| q_j - q_k \|).
\end{equation}
\twocolumngrid
The coordinates of $\Phi$ in $\R^m$ 
are then obtained by calculating the functions $G^2$ and $G^3$ for 
various values of the parameters $\eta$, $R_s$, $\lambda$, $\zeta$. 
One can easily check that $\Phi$ is invariant under rotation and 
permutation. Our goal is to study the relevance of this choice 
of fingerprint $\Phi$: Is this representation able to discriminate different 
structures ? What are the optimal parameters?  
How many functions should be used?

Our strategy is to estimate for various couples of configurations 
$C_i$, $C_j$ the distance in the space of descriptors $ \| \Phi(C_i) 
- \Phi(C_j) \|$, and the distance in the Cartesian space $\tilde{d}_2(C_i,C_j)$,
before computing their correlation. We will then be able 
to understand whether the representation is ambiguous or not: typically, a wrong 
representation would lead to high Atomic Configuration Distance $\tilde{d}_2$ (\acd) and low fingerprint difference,
or conversely low \acd{} and high fingerprint distance.

\begin{figure}[h]
  \centering
      \includegraphics[width=0.9\linewidth]{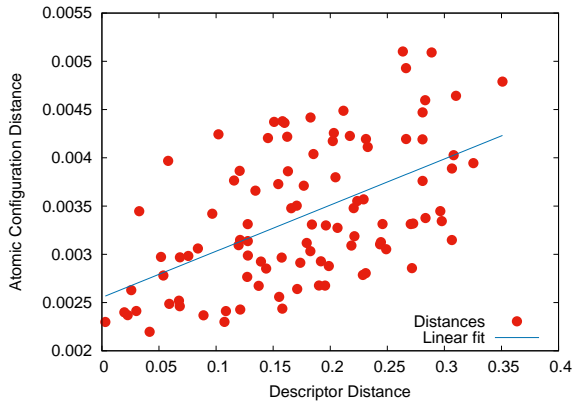}
      \caption{Correlation between \acd{} and descriptor distance 
        for two fingerprints, $\sigma=2$}
      \label{fig:correlationa}
\end{figure}

\begin{figure}[h]
  \centering
      \includegraphics[width=0.9\linewidth]{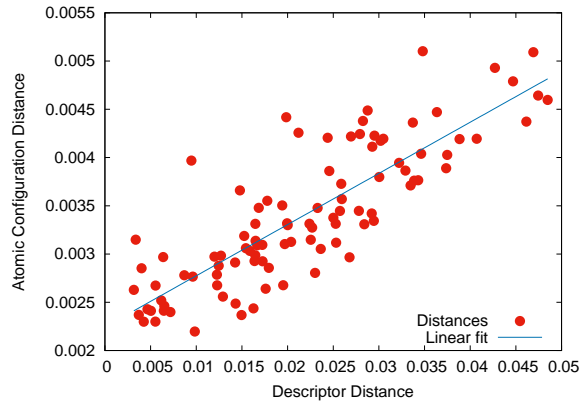}
      \caption{Correlation between \acd{} and descriptor distance 
        for a new set of fingerprints, $\sigma=2$}
      \label{fig:correlationb}
\end{figure}

We generate a set of configurations $C_i$ of a Lennard-Jones fluid
with a molecular dynamics simulation (see Appendix~\ref{sec:SM}).
As a first example, we apply the procedure by taking only one $G^2$
and one $G^3$ functions to represent the environment,
 with $\Rc=8.52$  \AA, $\eta=0.5$, $R_s=0$ \AA, $\lambda=-1$ and $\zeta=1$.
In Figure~\ref{fig:correlationa}, we observe the expected
correlation between the descriptor distance and the \acd{} for this
choice of descriptors but with an important dispersion. 
After a manual trial and error search, we found a set of 
functions giving a better correlation (see Appendix~\ref{sec:symm}), 
as displayed in Figure~\ref{fig:correlationb}. To quantify the increase of
faithfulness of the fingerprints, and given that the correlation seems linear, 
we perform the least square fit of the \acd{} as a function of the
descriptor distance, and compute for each case the associated correlation
coefficient $r\in[0,1]$. The first set of fingerprints has a correlation $r=0.55$
and the second $r=0.84$.

\begin{figure}[h]
  \centering
      \includegraphics[width=0.9\linewidth]{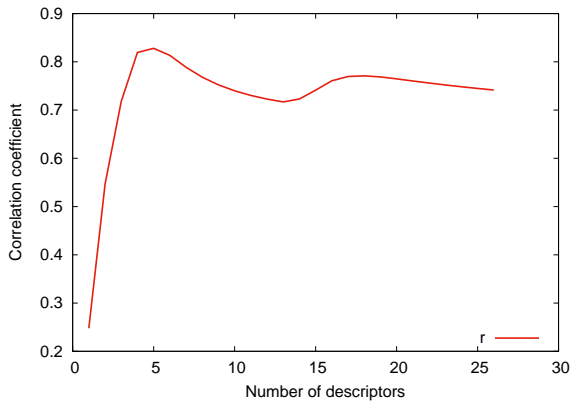}
      \caption{Evolution of the correlation coefficient $r$ when adding $G^2$ functions}
      \label{fig:bopt}
\end{figure}

Another point is to estimate the necessary and sufficient number of functions
to describe the configurations. Figure~\ref{fig:bopt}
shows  the evolution of the correlation coefficients $r$ 
when adding $G^2$ functions. We fix $\Rc=8.52$~\AA, $R_s=0$~\AA,
and consider additional values of $\eta$ obtained as
$\eta_1=1$, $\eta_2=\frac{\eta_1}{2}$, $\eta_3=\frac{\eta_2}{2}$, ...until $\eta_{13}=\frac{\eta_{12}}{2}$.. Then we choose the same values
of $\eta$ for $R_s=2.$ \AA. This
makes a total of 26 functions. We add the descriptors one by one, perform the
least square fit for each corresponding fingerprint and study the evolution of
the correlation coefficient $r$.
We first observe an increasing
correlation as expected, but after a critical number of functions,
$r$ decreases or remains stable. We also note an increase of the correlation when adding the first
symmetry functions with $R_s=2$ \AA. Other functions 
should be added, such as $G^3$ functions, to increase the representation capacity
of the set of descriptors. The decrease of correlation may seem surprising at first.
It is due to the fact that lower values of $\eta$ tend to generate functions
describing the environment far from the central particle. On the other hand, the
\acd{} has been implemented with a cut-off function as described in section~\ref{sec:frac}.
As a result, the description of the environment near the cut-off radius $\Rc$ is less correlated
to the \acd, hence the decreased correlation. This point is important in practice for several
applications. Typically for numerical potentials, using symmetry functions describing the
environment near the cut-off radius is not relevant since nearer neighbors have
a stronger influence on the potential.  

\begin{figure}[h]
  \centering
      \includegraphics[width=0.9\linewidth]{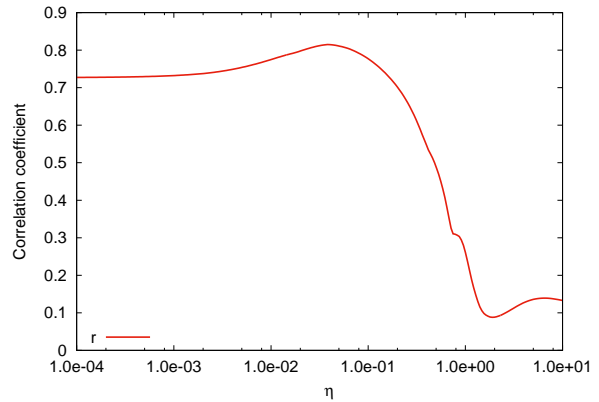}
      \caption{Evolution of correlation coefficient $r$ with $\eta$}
      \label{fig:etaR}
\end{figure}

\begin{figure}[h]
  \centering
      \includegraphics[width=0.9\linewidth]{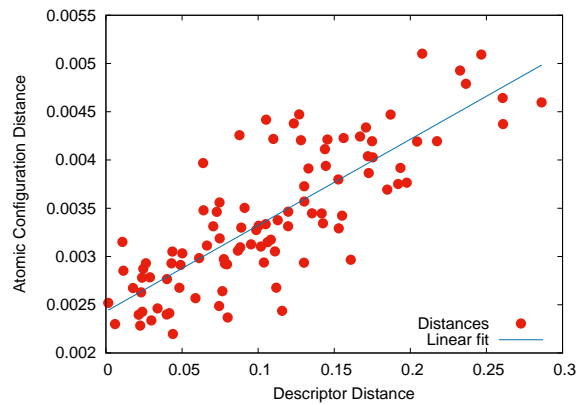}
      \caption{Correlation between \acd{} and descriptor distance 
        for one optimized $G^2$ function}
      \label{fig:correlationc}
\end{figure}

Let us now discuss how this procedure can be used to optimize the
set of parameters used for the descriptors. As a simple application, we aim at optimizing
the parameter $\eta$ in~(\ref{eq:G2}) for only one $G^2$ function. We
fix $R_s=0$~\AA, $\Rc=8.52$~\AA and study the correlation coefficient $r$ as a function of $\eta$ in Figure~\ref{fig:etaR}.
The graph shows a maximal value around $\eta^*=0.0381$, which
corresponds to a Gaussian of standard deviation~$3.6$. It can be understood as a typical length to describe the system around its central atom. 
The correlation for low values of $\sigma$ is still correct, but it drastically decreases when
$\eta$ becomes too large. This can be explained by the fact that the variance of the Gaussian tends to zero
when $\eta$ becomes large, so the $G^2$ function takes values close to
$0$ whatever the environment, and does not describe the environment
anymore. The correlation graph associated with the optimal value $\eta^*$ is displayed in 
Figure~\ref{fig:correlationc}. Even though the correlation is good, we observe as in Figure~\ref{fig:correlationa}
that for some couples of configurations, the descriptor distance is $0$ whereas the \acd{} is not, which is
a crucial problem. It suggests that our descriptor should also maximize the quantity 
\[
\underset{i,j}{\min}\ \| \Phi(C_i) - \Phi(C_j) \|. 
\] 
Several tracks can therefore be followed when $C_i$ and $C_j$ do not belong to
the same configuration to
optimize a fingerprint. First, a criterium to optimize should be chosen (maximizing the 
correlation or minimizing the lack of injectivity of~$\Phi$). As a second
step, the set of symmetry functions can be optimized either by a greedy technique (adding the functions
one by one with associated optimized parameters) or by defining a set of functions and optimizing
all parameters together. The first option does not provide the optimal set of functions,
but is easier to implement. The second one possibly requires to explore a large set of
parameters, and more advanced optimization methods should be investigated.

To summarize, the \acd{} is a way to quantify the conservation
of information when using a descriptor. It can be used to
indicate if functions are relevant to represent a given set of
structures. It also allows to quantify the improvement in the
representation obtained by adding a new describing function, and to optimize
the parameters in the descriptors. 
This is particularly
important in the context of Machine Learning, since the dimension of the 
fingerprint $m$ (\textit{i.e.} the number of functions in the representation)
is the dimension of the interpolation space for the methods (Neural
Networks or Kernel methods). Reducing the size of the input space
$\R^m$ is of crucial importance to increase the efficiency of this
interpolation (or learning) procedure. 
Lower values of $m$ also 
allow to decrease the computational cost of
evaluating the numerical potential, which is significant in order
to perform faster Molecular Dynamics simulations.

\subsection{Structural Analysis}

\begin{figure}[h]
  \centering
      \includegraphics[width=0.9\linewidth]{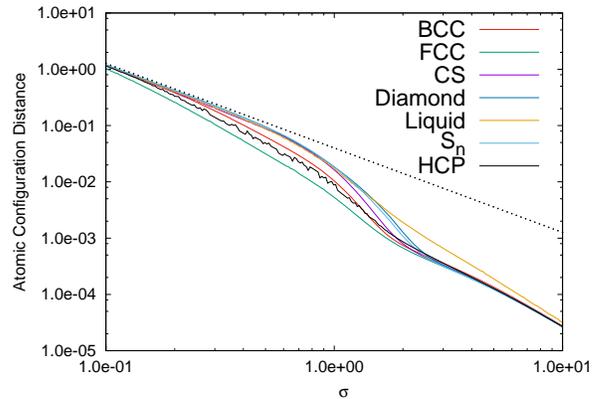}
      \caption{Atomic Configuration Distance to each reference structure as a function of $\sigma$}
      \label{fig:StruSigma}
\end{figure}

\begin{figure}[h]
  \centering
      \includegraphics[width=0.9\linewidth]{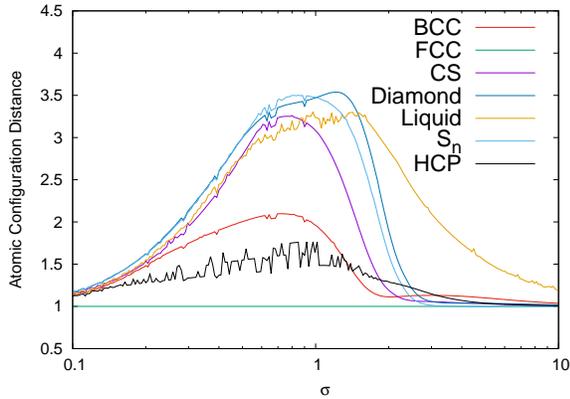}
      \caption{Atomic Configuration Distance to each reference
        structure normalized by the 
        distance to FCC, as a function of $\sigma$}
      \label{fig:StruSigmaRef}
\end{figure}

The goal of this section is to demonstrate in a simple case
that our method  can be used to identify
the local structure of a material. We restrict ourselves to
the structural analysis of a crystal with one chemical
element, but extensions to crystals with several elements or even molecules
are straightforward.
For the testing procedure, we gather
a database of crystalline structures $C_i$ at $0$ Kelvin (at a given density, see Appendix~\ref{sec:SM} for more details on the generation of this database):

\begin{itemize}
\item Body Centered Cubic (BCC),
\item Face Centered Cubic (FCC),
\item Simple Cubic (CS),
\item Hexagonal Close Packed (HCP)
\item diamond,
\item liquid,
\item Sn$\beta$ ($\beta$ structure of tin).
\end{itemize} 
We next compare a structure $C$ of FCC at 100 Kelvins obtained by
molecular dynamics simulation 
to each element of the database $C_i$, 
by computing the distance in the fingerprints space
$
\| \Phi(C) - \Phi(C_i) \|
$
and \acd{} $\tilde{d}_2(C,C_i)$. We use the fingerprint described in Appendix~\ref{sec:symm}. 
Figure~\ref{fig:Stru1} represents 
  $\tilde{d}^2(C,C_i) $ as a function of $ \| \Phi(C) - \Phi(C_i)\|$
for each reference structure $C_i$. 
For both \acd{} and fingerprints,
 the structure at 100 Kelvin is closer to FCC than
to any other structure. This shows that \acd{} is efficient
in recognizing the local structure of a material, and it also validates
the choice of fingerprint used for the recognition of these environments. 
In order to better understand the influence of the smearing parameter $\sigma$,
we display in Figure~\ref{fig:StruSigma}
the evolution of the Atomic Configuration Distance as a function of~$\sigma$.
The numerical
values vary but the conclusion remains unchanged: the distance to the FCC structure
is always the lowest. When $\sigma$
decreases, the accuracy of the distance increases, so small variations
of the structure due to thermal fluctuations tend to be understood as a 
structural difference. On the other hand, when $\sigma$ is large, the precision
is very low (or equivalently, the representing densities are very smooth), and all
the configurations tend to be recognized as equivalent. The asymptotic behavior when
$\sigma$ tends to 0 is also displayed; the distance scales as $\sigma^{- \sfrac{3}{2}}$, as predicted by the analysis
performed in Appendix~\ref{sec:asympt}. The same behavior is observed for
larger values of $\sigma$. Lastly, this distance induces a classification of similarity between structures. Indeed,
Figure~\ref{fig:StruSigmaRef} shows the distance to each reference structure over the
distance to FCC. All distances are larger than the distance to FCC, 
but with an ordering. Indeed,
the distance to HCP and BCC is higher than the distance to FCC typically by a factor 1.5
or 2. On the other hand, distances to the other elements up to 4 times higher
than the distance to FCC. This can be explained by  the chemical similarity of HCP and BCC 
with the FCC structure. The chosen descriptor fails to provide this detailed
study, since its distance to HCP is the highest one.
\begin{figure}[h]
  \centering
      \includegraphics[width=0.9\linewidth]{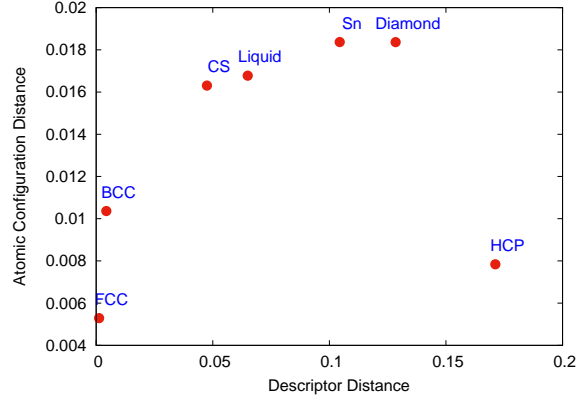}
      \caption{Distance from structure $C$ to reference structures, $\sigma=1$}
      \label{fig:Stru1}
\end{figure}

\section{Conclusion}

Our goal in this work was to compare atomic configurations for various
purposes, such as structure recognition or Machine Learning
methods.  To this end, we introduced a formal description 
of what we call
\emph{atomic environment} and \emph{atomic configuration},
which provides a firm mathematical ground for defining
a distance between configurations. Then, we showed that 
an environment can be represented as a permutation
invariant density of probability, the \fracc. This
consideration naturally led us to define a permutation
invariant distance between atomic environments of an
arbitrary number of atoms $n_1$ and $n_2$. In particular, the Gaussian case
provides an analytical distance whose computational cost
 scales as $n_1 n_2$. We showed that
taking the minimum over rotations of this distance creates
a permutation and rotation invariant Atomic Configuration Distance (\acd). 
This means that a distance can
be defined between two configurations. The significant difference
with previous methodologies (notably \rmsd) is that
the \acd{}  is intrinsically permutation invariant, so that
performing the minimum over permutations is unnecessary, and comparisons
of environments with hundreds of atoms become tractable. This
is of paramount importance for condensed matter systems.

Many applications can be envisioned. We considered two
of them as an example. First, we studied the capability of
fingerprints to reproduce the configurational information 
of a structure. This is of critical importance in several fields,
in particular Machine Learning methods for numerical potentials. We showed that the 
introduced distance does not answer the issue
of the choice of the descriptors by itself, but it provides a quantitative
assessment of the improvement in the description provided by an additional 
representing function. This allows typically to implement greedy techniques, where 
parameters of a new function are optimized in order to obtain the maximal
correlation $r$ between the \acd{} and the descriptor's distance.
A second application is structural analysis.
By comparing a local structure to a reference data set of structures
at 0 Kelvin, we showed that the \acd{}~(\ref{eq:metric}) is able to recognize 
the structure of a material at a positive temperature. 
Let us however mention that our methodology is not a
way to characterize a structure from scratch. If the studied structure
is not in the database, it will not be recognized.
 
Finally, we want to emphasize that we described the presented tools
in a quite general and abstract way to allow for various applications.
At several points of the derivation of the distances, 
choices can be made (such as the functions for representing
the presence of an atom in the \fracc{} or the norm to compare densities).
Moreover, other applications can be considered: comparing molecules, testing other
fingerprints, or using directly
the similarity $S$ defined in~(\ref{eq:S}) as a Kernel, as for the \soap{}\cite{Bartok2013repr},
(but taking the minimum over rotations rather than integrating over them and using the analytical
expression~(\ref{eq:S})). 
Another important issue is the sparsification,
\textit{i.e.} the construction of a database minimizing the redundancy of information. Indeed,
in the context of numerical potentials, it is crucial to have the smallest possible database 
to decrease the computation cost of the potential.

\appendix

\section{Proof that ACD is a Distance}
\label{appendix:A}
We here prove that the application defined in~(\ref{eq:metric}) is
a distance on $\Ct$.
We remind that to be a distance on $\Ct$, $\tilde{d_2}$ must verify,
for any configurations $\widetilde{C}_1$, $\widetilde{C}_2$,
$\widetilde{C}$$\in\Ct$, the following properties:
\begin{itemize}
\item $\tilde{d_2}(\widetilde{C}_1,\widetilde{C}_2)
      =\tilde{d_2}(\widetilde{C}_2,\widetilde{C}_1)$,
\item $\tilde{d_2}(\widetilde{C}_1,\widetilde{C}_2)=0$ if and
only if $\widetilde{C}_1=\widetilde{C}_2$,
\item $\tilde{d_2}(\widetilde{C}_1,\widetilde{C}_2)\leq 
\tilde{d_2}(\widetilde{C}_1,\widetilde{C}) +
\tilde{d_2}(\widetilde{C},\widetilde{C}_2)$.
\end{itemize} 
First, $d_2$ being invariant by permutation, and since the rotation 
can be taken over $C_1$ or $C_2$ in~(\ref{eq:metric}),
the definition of the distance is easily seen to be independent 
of the choice of the representing environments 
$C_1\in \widetilde{C}_1$ and $C_2\in \widetilde{C}_2$.  
Secondly, the optimization problem always has a solution since the space of rotations
$\mathcal{R}$ is compact and the application $R\mapsto d_2\,(C_1,RC_2)$ is continuous.
Therefore, the application is well-defined on $\Ct \times \Ct$. 
Now, the triangle inequality results from the isomorphism with the 
associated densities $\rho_1$, $\rho_2$. More precisely, for another configuration 
$\widetilde{C}\in\Ct$, with a representing element $C\in \widetilde{C}$ 
and associated density $\rho$, 
\[
\|\rho_1 - \rho_2\|_{L^2} \leq \|\rho_1 - \rho\|_{L^2} + \|\rho - \rho_2\|_{L^2},
\]
So, for any rotation $R$,
\[
d_2(C_1,RC_2) \leq d_2(C_1,C) + d_2(C,RC_2).
\]
Moreover, given that $C\in\widetilde{C}$ is arbitrary, 
we can choose it such that $d_2(C_1,C)= \underset{C'\in \widetilde{C}} 
{\inf} d_2(C_1,C') $, which leads to
\[
d_2(C_1,RC_2) \leq \tilde{d_2}(\widetilde{C}_1,\widetilde{C}) + d_2(C,RC_2).
\]
Now, by taking the infimum over rotations $R$, first on the left hand side, 
and then on the right side, we obtain the triangle 
inequality
\[
\tilde{d_2}(\widetilde{C}_1,\widetilde{C}_2) \leq 
\tilde{d_2}(\widetilde{C}_1,\widetilde{C}) + \tilde{d_2}
(\widetilde{C},\widetilde{C}_2).
\]

The last point is to prove that $\widetilde{C}_1=\widetilde{C}_2$ if and only if 
$\tilde{d_2}(\widetilde{C}_1,\widetilde{C}_2)=0$. First, it is clear that 
$\tilde{d_2}(\widetilde{C},\widetilde{C})=0$ for any $\widetilde{C}\in \Ct$. Now, 
take $\widetilde{C}_1$, $\widetilde{C}_2\in \widetilde{\mathcal{C}}$ and $C_1\in 
\widetilde{C}_1$, $C_2\in \widetilde{C}_2$ such that $ \tilde{d_2}(\widetilde{C}_1,
\widetilde{C}_2)=0$. This implies
\[
\underset{R\in\mathcal{R}}{\inf}\, d_2\,(C_1,RC_2)=0.
\]
Therefore, there exists $R_0\in \mathcal{R}$ such that $ d_2\,(C_1,R_0 C_2)=0$, and 
given that $d_2$ is a distance up to a permutation, there exists $\sigma_0\in 
\mathcal{S}$ such that $C_1=\sigma_0 R_0 C_2$, \textit{i.e.} $C_1$ and $C_2$ belong to 
the same class and $\widetilde{C}_1=\widetilde{C}_2$.

\smallskip

\onecolumngrid
\section{Analytical Computation of the overlap matrices in the Gaussian
 case}
\label{appendix:B}

We prove formulas~(\ref{eq:S}) and~(\ref{eq:ERMSD}). First,
\[
S(\rho_1,\rho_2)=\int_{\R^3} \rho_1 \rho_2  = \frac{1}{n_1 n_2} \, \sum_{i=1}^{n_1} 
\sum_{j=1}^{n_2} \, \int_{\R^3}\varphi_{q_i}^{\sigma}(q)\, \varphi_{q_j'}^{\sigma}(q) \, \dd q.
\]
Any integral in the double sum can be analytically computed as

\renewcommand{\arraystretch}{2.5}
\[
\begin{array}{*2{>{\displaystyle}l}}
\int_{\R^3}\varphi_{q_i}^{\sigma}(q)\, \varphi_{q_j'}^{\sigma}(q) \, \dd q & = \int_{\R^3} 
\frac{1}{(2\pi\sigma^2)^3} \exp \left( -\frac{1}{2\sigma^2} \Big[ (q-q_i)^2 
+(q - q_j')^2  \Big] \right) \dd q \\
& =\int_{\R^3} \frac{1}{(2\pi\sigma^2)^3} \exp \left( -\frac{1}{2\sigma^2} 
\left( 2q^2 -2 q \cdot q_i - 2 q\cdot q_j' + q_i^2 +q_j'^2  \right) \right) \dd q \\
& =\int_{\R^3} \frac{1}{(2\pi\sigma^2)^3} \exp \left( -\frac{1}{\sigma^2} \left( 
\left( q - \frac{q_i + q_j'}{2}\right)^2 - \frac{( q_i + q_j')^2}{4} + 
\frac{q_i^2+q_j'^2}{2} \right) \right) \dd q \\
& = \frac{1}{(2\pi\sigma^2)^3} \left[  \int_{\R^3} \exp \left( -\frac{1}{\sigma^2} 
\left( q - \frac{q_i + q_j'}{2}\right)^2 \right) \dd q \right] \, \exp \left( 
-\frac{1}{4\sigma^2} \left( q_i -q_j' \right)^2 \right). 
\end{array}
\]
Finally, the identity
 \[ \int_{\R^3} \exp \left( -\frac{1}{\sigma^2} \left( q - \frac{q_i + q_j'}{2}\right)^2 \right) 
\dd q= (\sigma^2 \pi)^{\sfrac{3}{2}}\] gives the desired formula.

\twocolumngrid

\section{Asymptotics of \acd}
\label{sec:asympt}

We study the asymptotic behavior of formula~(\ref{eq:S}) when
$\sigma\to 0$ and $\sigma\to +\infty$. We rewrite the overlap factor as
\[
S(\sigma) = \frac{ (\pi \sigma^2)^{- \frac{3}{2} }}
{8 n_1 n_2} \sum_{i=1}^{n_1} 
\sum_{j=1}^{n_2} \exp\left(- \frac{(q_i - q'_j)^2}{4\sigma^2} \right),
\]
where $S$ is considered as a function of $\sigma$ and 
the two environments $C_1=(q_i)_{i=1}^{n_1}$, 
$C_2=(q'_i)_{i=1}^{n_2}\in\C$ are fixed. When $\sigma\to +\infty$,
it holds for all $i$, $j$,
\[
\exp\left(- \frac{(q_i - q'_j)^2}{4\sigma^2} \right)
\xrightarrow[\sigma \to +\infty]{} 1.
\]
As a result, 
\[
 S(\sigma)  \underset{+\infty}
{\sim} \frac{ (\pi \sigma^2)^{- \frac{3}{2} }}
{8n_1 n_2}.
\]
Given that distances are obtained by taking the square root of overlap factors, the \acd{} scales as
$\sigma^{- \sfrac{3}{2} }$ when $\sigma\to +\infty$. 

In the case $\sigma\to 0$, one has to distinghish between the cases $q_i \neq q_j'$, for 
which $\exp\left(- \frac{(q_i - q_j)^2}{4\sigma^2} \right)$ converges to~0 very fast; and $q_i = q_j'$, in which 
case this factor is always equal to~1. Since in the definition of the \acd{} distance~\eqref{eq:l2} there are two
overlap factors involving the same density, the dominant terms in~\eqref{eq:l2} are the elements in the double sums
$S(\rho_1,\rho_1)$ and $S(\rho_2,\rho_2)$ for which $i=j$, and they are equal to~1. The asymptotic behavior is then determined by the diverging prefactor, which shows that the \acd{} scales as $\sigma^{- \sfrac{3}{2} }$ when $\sigma\to 0$. 

\section{Simulated Annealing}
\label{sec:siman}
We shortly describe here the simulated annealing method\cite{SA1987,SA2005}
used to solve the optimization problem in~(\ref{eq:metric}).
We use a Metropolis Random Walk with Gaussian proposals\cite{MRRTT53}, together with
a schedule for temperature decrease. We also decrease the variance of
the proposals in order to maintain the acceptance rate roughly constant.
We denote $\theta=(\alpha,\beta,\gamma)$ the set of angles parametrizing
the rotation. We choose a 
starting point $\theta_0$, $\xi_0>0$, $\lambda_0>0$, a decrease rate
$\tau$. The iteration proceeds as follows for each $k\geq 1$:
\begin{itemize}
\item define $\tilde{\theta}_{k+1}=\theta_k +\xi_k G_k$, where $G_k\sim \mathcal{N}(0,I_d)$
is a standard 3-dimensional Gaussian vector with identity covariance,
\item compute 
\begin{equation}
\renewcommand{\arraystretch}{1.5}
\begin{array}{*2{>{\displaystyle}l}}
\dps a(\tilde{\theta}_{k+1}, \theta_k )=\exp \left( - 
\frac{1}{\lambda_k}[ J(\tilde{\theta}_{k+1}) - J(\theta_k) ] \right) \nonumber
\end{array}
\end{equation}
and $r(\tilde{\theta}_{k+1})=\min \left ( 1, a(\tilde{\theta}_{k+1})
\right )$,
\item accept the move with a probability $r(\tilde{\theta}_{k+1})$ and set
$\theta_{k+1}=\tilde{\theta}_{k+1}$ in this case; 
otherwise reject the move and set $\theta_{k+1}=\theta_k$,
\item set $\lambda_{k+1}=\tau\lambda_k$ and $\xi_{k+1}=\tau \xi_k$.
\end{itemize}
We fix the number of iterations (typically 1000 steps), and run the algorithm from various 
starting points $\theta_0$ on a uniform grid, typically 2 or 3 points in each
direction. We use $\tau=0.99$. The initial values $\xi_0$ and
$\lambda_0$ are choosen such that the acceptance rate remains of order $60-70\%$ at each stage. 
Typical values are $\xi_0=1$ and $\lambda_0=10^{-4}$. The value of $\lambda_0$ is derived from
the following computation \[ J(\tilde{\theta}_{1}) - J(\theta_0) 
\simeq \xi_0 G_0 \cdot  \nabla J(\theta_0),
\]
which suggests to choose $\lambda_0 \sim \xi_0 \| \nabla J(\theta_0)\|$. 
Another option could be to use a single starting point but perform sequences of heating-cooling,
\textit{i.e.} increasing-decreasing $\lambda_k$ and $\xi_k$ to explore the different
parts of the domain.

\section{Symmetry functions}
\label{sec:symm}
We fix the values of the parameters $\eta=$0.3 and $R_{\rm cut}$=8.52~{\AA} for the set of $G^2$ symmetry functions used
as a descriptor in the applications presented in Section~\ref{sec:appli} (correlation graph and
structural analysis). The remaining $R_s$ parameters take the values
{1.5, 2.0, 2.2, 2.4, 2.6, 2.8, 3.0, 3.5} \AA. These functions correspond to Gaussian functions
centered at different radiuses to represent the radial information
of a configuration.

\section{Generation of databases}
\label{sec:SM}

We give here precisions on the generation of the databases used for the
numerical applications. First, we used in section IV-A a database given by MD
simulations of a Lennard-Jones fluid. Secondly, a reference database
of structure was generated for the structural analysis performed in
section IV-B, and another MD simulation
was realized to compute a configuration of CFC at 100 Kelvin.

\subsection{Fluid}

The configurations of fluid were generated through a MD simulation of 
a Lennard-Jones NVT system at 1000 Kelvin. Langevin equations are used
with a truncated and shifted Lennard-Jones potential $\vlj$. The
values of the parameters are $\varepsilon=1.6567944\times
10^{-21}\mathrm{J}$ and $\sigma=3.405$ \AA, reproducing
thermodynamic properties of Argon. A cut-off radius $\Rc=2.5 \sigma$
was employed. The timestep for integration is equal to $5 \times 10^{-15}$ s,
and the dynamics is integrated over $20 000$ time steps.

Finally, the configurations are obtained by recentering the
environment on each atom and using
the periodic conditions to define its neighbors.

\subsection{Reference Database of Structures}

The environments describing crystals (CFC, BCC, etc) at 0 Kelvin
are well known. The issue is to use structures with
a correct density, since comparing structures at different densities
is irrelevant in our case. Indeed, identical structures at different
densities will be considered as different by the \acd. 
As a result, we use CFC as a reference structure, and its 
lattice constant is chosen such that the structure has a minimal
energy, or equivalently a zero pressure. The equilibrium density
is $1.8025$ kg.m$^{-3}$. The other structures are
generated with the same density (lattice parameters are adjusted since
the shape of the unit cell and the number of atoms per unit cell change).
For a CFC structure at 100 Kelvin, the same MD simulation is run over
20 000 steps, but at 100 Kelvin.

Note that in addition to the invariance by permutation
and rotation, we could here add an invariance by dilatation of an
environment. As a result we could search for the dilatation that
best allows to match two environments. Nevertheless, this would add a degree of search (and therefore increase the computational cost) whereas the physical approach of keeping the density fixed is straightforward and inexpensive.


\begin{thebibliography}{10}%
\makeatletter
\providecommand \@ifxundefined [1]{%
 \ifx #1\undefined \expandafter \@firstoftwo
 \else \expandafter \@secondoftwo
\fi
}%
\providecommand \@ifnum [1]{%
 \ifnum #1\expandafter \@firstoftwo
 \else \expandafter \@secondoftwo
\fi
}%
\providecommand \enquote [1]{``#1''}%
\providecommand \bibnamefont  [1]{#1}%
\providecommand \bibfnamefont [1]{#1}%
\providecommand \citenamefont [1]{#1}%
\providecommand\href[0]{\@sanitize\@href}%
\providecommand\@href[1]{\endgroup\@@startlink{#1}\endgroup\@@href}%
\providecommand\@@href[1]{#1\@@endlink}%
\providecommand \@sanitize [0]{\begingroup\catcode`\&12\catcode`\#12\relax}%
\@ifxundefined \pdfoutput {\@firstoftwo}{%
 \@ifnum{\z@=\pdfoutput}{\@firstoftwo}{\@secondoftwo}%
}{%
 \providecommand\@@startlink[1]{\leavevmode}%
 \providecommand\@@endlink[0]{}%
}{%
 \providecommand\@@startlink[1]{%
  \leavevmode
  \pdfstartlink
   attr{/Border[0 0 1 ]/H/I/C[0 1 1]}%
   user{/Subtype/Link/A<</Type/Action/S/URI/URI(#1)>>}%
  \relax
 }%
 \providecommand\@@endlink[0]{\pdfendlink}%
}%
\providecommand \url  [0]{\begingroup\@sanitize \@url }%
\providecommand \@url [1]{\endgroup\@href {#1}{\urlprefix}}%
\providecommand \urlprefix [0]{URL }%
\providecommand \Eprint[0]{\href }%
\@ifxundefined \urlstyle {%
  \providecommand \doi [1]{doi:\discretionary{}{}{}#1}%
}{%
  \providecommand \doi [0]{doi:\discretionary{}{}{}\begingroup
  \urlstyle{rm}\Url }%
}%
\providecommand \doibase [0]{http://dx.doi.org/}%
\providecommand \Doi[1]{\href{\doibase#1}}%
\providecommand \bibAnnote [3]{%
  \BibitemShut{#1}%
  \begin{quotation}\noindent
    \textsc{Key:}\ #2\\\textsc{Annotation:}\ #3%
  \end{quotation}%
}%
\providecommand \bibAnnoteFile [2]{%
  \IfFileExists{#2}{\bibAnnote {#1} {#2} {\input{#2}}}{}%
}%
\providecommand \typeout [0]{\immediate \write \m@ne }%
\providecommand \selectlanguage [0]{\@gobble}%
\providecommand \bibinfo [0]{\@secondoftwo}%
\providecommand \bibfield [0]{\@secondoftwo}%
\providecommand \translation [1]{[#1]}%
\providecommand \BibitemOpen[0]{}%
\providecommand \bibitemStop [0]{}%
\providecommand \bibitemNoStop [0]{.\EOS\space}%
\providecommand \EOS [0]{\spacefactor3000\relax}%
\providecommand \BibitemShut [1]{\csname bibitem#1\endcsname}%
\bibitem{Karakoc15072006}%
  \BibitemOpen
  \bibfield{author}{%
  \bibinfo {author} {\bibfnamefont{E.}~\bibnamefont{Karakoc}}, \bibinfo
  {author} {\bibfnamefont{A.}~\bibnamefont{Cherkasov}},\ and\ \bibinfo {author}
  {\bibfnamefont{S.~C.}\ \bibnamefont{Sahinalp}},\ }%
  \bibfield{journal}{%
  \bibinfo {journal} {Bioinformatics}\ }%
  \textbf{\bibinfo {volume} {22}},\ \bibinfo {pages} {243} (\bibinfo {year}
  {2006})%
  \bibAnnoteFile{NoStop}{Karakoc15072006}%
\bibitem{hopping2004}%
  \BibitemOpen
  \bibfield{author}{%
  \bibinfo {author} {\bibfnamefont{S.}~\bibnamefont{Goedecker}},\ }%
  \bibfield{journal}{%
  \bibinfo {journal} {J.~ Chem. Phys.}\ }%
  \textbf{\bibinfo {volume} {120}},\ \bibinfo {pages} {9911} (\bibinfo {year}
  {2004})%
  \bibAnnoteFile{NoStop}{hopping2004}%
\bibitem{hopping2010}%
  \BibitemOpen
  \bibfield{author}{%
  \bibinfo {author} {\bibfnamefont{M.}~\bibnamefont{Amsler}}\ and\ \bibinfo
  {author} {\bibfnamefont{S.}~\bibnamefont{Goedecker}},\ }%
  \bibfield{journal}{%
  \bibinfo {journal} {J.~ Chem. Phys.}\ }%
  \textbf{\bibinfo {volume} {133}},\ \bibinfo {eid} {224104} (\bibinfo {year}
  {2010})%
  \bibAnnoteFile{NoStop}{hopping2010}%
\bibitem{schutt2014}%
  \BibitemOpen
  \bibfield{author}{%
  \bibinfo {author} {\bibfnamefont{K.~T.}\ \bibnamefont{Sch\"utt}}, \bibinfo
  {author} {\bibfnamefont{H.}~\bibnamefont{Glawe}}, \bibinfo {author}
  {\bibfnamefont{F.}~\bibnamefont{Brockherde}}, \bibinfo {author}
  {\bibfnamefont{A.}~\bibnamefont{Sanna}}, \bibinfo {author}
  {\bibfnamefont{K.~R.}\ \bibnamefont{M\"uller}},\ and\ \bibinfo {author}
  {\bibfnamefont{E.~K.~U.}\ \bibnamefont{Gross}},\ }%
  \bibfield{journal}{%
  \bibinfo {journal} {Phys. Rev. B}\ }%
  \textbf{\bibinfo {volume} {89}},\ \bibinfo {pages} {205118} (\bibinfo {year}
  {2014})%
  \bibAnnoteFile{NoStop}{schutt2014}%
\bibitem{SadeghiMetric2013}%
  \BibitemOpen
  \bibfield{author}{%
  \bibinfo {author} {\bibfnamefont{A.}~\bibnamefont{Sadeghi}}, \bibinfo
  {author} {\bibfnamefont{S.~A.}\ \bibnamefont{Ghasemi}}, \bibinfo {author}
  {\bibfnamefont{B.}~\bibnamefont{Schaefer}}, \bibinfo {author}
  {\bibfnamefont{S.}~\bibnamefont{Mohr}}, \bibinfo {author}
  {\bibfnamefont{M.~A.}\ \bibnamefont{Lill}},\ and\ \bibinfo {author}
  {\bibfnamefont{S.}~\bibnamefont{Goedecker}},\ }%
  \bibfield{journal}{%
  \bibinfo {journal} {J.~ Chem. Phys.}\ }%
  \textbf{\bibinfo {volume} {139}},\ \bibinfo {pages} {184118} (\bibinfo {year}
  {2013})%
  \bibAnnoteFile{NoStop}{SadeghiMetric2013}%
\bibitem{ogano2009}%
  \BibitemOpen
  \bibfield{author}{%
  \bibinfo {author} {\bibfnamefont{A.~R.}\ \bibnamefont{Oganov}}\ and\ \bibinfo
  {author} {\bibfnamefont{M.}~\bibnamefont{Valle}},\ }%
  \bibfield{journal}{%
  \bibinfo {journal} {J.~ Chem. Phys.}\ }%
  \textbf{\bibinfo {volume} {130}},\ \bibinfo {eid} {104504} (\bibinfo {year}
  {2009})%
  \bibAnnoteFile{NoStop}{ogano2009}%
\bibitem{GAPAnderson}%
  \BibitemOpen
  \bibfield{author}{%
  \bibinfo {author} {\bibfnamefont{L.-F.}\ \bibnamefont{Arsenault}}, \bibinfo
  {author} {\bibfnamefont{A.}~\bibnamefont{Lopez-Bezanilla}}, \bibinfo {author}
  {\bibfnamefont{O.~A.}\ \bibnamefont{von Lilienfeld}},\ and\ \bibinfo {author}
  {\bibfnamefont{A.~J.}\ \bibnamefont{Millis}},\ }%
  \bibfield{journal}{%
  \bibinfo {journal} {Phys. Rev. B}\ }%
  \textbf{\bibinfo {volume} {90}},\ \bibinfo {pages} {155136} (\bibinfo {year}
  {2014})%
  \bibAnnoteFile{NoStop}{GAPAnderson}%
\bibitem{Molecular2013}%
  \BibitemOpen
  \bibfield{author}{%
  \bibinfo {author} {\bibfnamefont{K.}~\bibnamefont{Hansen}}, \bibinfo {author}
  {\bibfnamefont{G.}~\bibnamefont{Montavon}}, \bibinfo {author}
  {\bibfnamefont{F.}~\bibnamefont{Biegler}}, \bibinfo {author}
  {\bibfnamefont{S.}~\bibnamefont{Fazli}}, \bibinfo {author}
  {\bibfnamefont{M.}~\bibnamefont{Rupp}}, \bibinfo {author}
  {\bibfnamefont{M.}~\bibnamefont{Scheffler}}, \bibinfo {author}
  {\bibfnamefont{O.~A.}\ \bibnamefont{von Lilienfeld}}, \bibinfo {author}
  {\bibfnamefont{A.}~\bibnamefont{Tkatchenko}},\ and\ \bibinfo {author}
  {\bibfnamefont{K.-R.}\ \bibnamefont{Müller}},\ }%
  \bibfield{journal}{%
  \bibinfo {journal} {J.~ Chem. Theory Comput.}\ }%
  \textbf{\bibinfo {volume} {9}},\ \bibinfo {pages} {3404} (\bibinfo {year}
  {2013})%
  \bibAnnoteFile{NoStop}{Molecular2013}%
\bibitem{seko2014}%
  \BibitemOpen
  \bibfield{author}{%
  \bibinfo {author} {\bibfnamefont{A.}~\bibnamefont{Seko}}, \bibinfo {author}
  {\bibfnamefont{A.}~\bibnamefont{Takahashi}},\ and\ \bibinfo {author}
  {\bibfnamefont{I.}~\bibnamefont{Tanaka}},\ }%
  \bibfield{journal}{%
  \bibinfo {journal} {Phys. Rev. B}\ }%
  \textbf{\bibinfo {volume} {90}},\ \bibinfo {pages} {024101} (\bibinfo {year}
  {2014})%
  \bibAnnoteFile{NoStop}{seko2014}%
\bibitem{GAPTungsten}%
  \BibitemOpen
  \bibfield{author}{%
  \bibinfo {author} {\bibfnamefont{W.~J.}\ \bibnamefont{Szlachta}}, \bibinfo
  {author} {\bibfnamefont{A.~P.}\ \bibnamefont{Bart\'ok}},\ and\ \bibinfo
  {author} {\bibfnamefont{G.}~\bibnamefont{Cs\'anyi}},\ }%
  \bibfield{journal}{%
  \bibinfo {journal} {Phys. Rev. B}\ }%
  \textbf{\bibinfo {volume} {90}},\ \bibinfo {pages} {104108} (\bibinfo {year}
  {2014})%
  \bibAnnoteFile{NoStop}{GAPTungsten}%
\bibitem{behler2014ref}%
  \BibitemOpen
  \bibfield{author}{%
  \bibinfo {author} {\bibfnamefont{J.}~\bibnamefont{Behler}},\ }%
  \bibfield{journal}{%
  \bibinfo {journal} {J.~ Phys.: Condens. Matter}\ }%
  \textbf{\bibinfo {volume} {26}},\ \bibinfo {pages} {183001} (\bibinfo {year}
  {2014})%
  \bibAnnoteFile{NoStop}{behler2014ref}%
\bibitem{handley2014next}%
  \BibitemOpen
  \bibfield{author}{%
  \bibinfo {author} {\bibfnamefont{C.~M.}\ \bibnamefont{Handley}}\ and\
  \bibinfo {author} {\bibfnamefont{J.}~\bibnamefont{Behler}},\ }%
  \bibfield{journal}{%
  \bibinfo {journal} {Eur. Phys. J.~ B}\ }%
  \textbf{\bibinfo {volume} {87}} (\bibinfo {year} {2014})%
  \bibAnnoteFile{NoStop}{handley2014next}%
\bibitem{Bartok2010GAP}%
  \BibitemOpen
  \bibfield{author}{%
  \bibinfo {author} {\bibfnamefont{A.~P.}\ \bibnamefont{Bart\'ok}}, \bibinfo
  {author} {\bibfnamefont{M.~C.}\ \bibnamefont{Payne}}, \bibinfo {author}
  {\bibfnamefont{R.}~\bibnamefont{Kondor}},\ and\ \bibinfo {author}
  {\bibfnamefont{G.}~\bibnamefont{Cs\'anyi}},\ }%
  \bibfield{journal}{%
  \bibinfo {journal} {Phys. Rev. Lett.}\ }%
  \textbf{\bibinfo {volume} {104}},\ \bibinfo {pages} {136403} (\bibinfo {year}
  {2010})%
  \bibAnnoteFile{NoStop}{Bartok2010GAP}%
\bibitem{Bartok2013repr}%
  \BibitemOpen
  \bibfield{author}{%
  \bibinfo {author} {\bibfnamefont{A.~P.}\ \bibnamefont{Bart\'ok}}, \bibinfo
  {author} {\bibfnamefont{R.}~\bibnamefont{Kondor}},\ and\ \bibinfo {author}
  {\bibfnamefont{G.}~\bibnamefont{Cs\'anyi}},\ }%
  \bibfield{journal}{%
  \bibinfo {journal} {Phys. Rev. B}\ }%
  \textbf{\bibinfo {volume} {87}},\ \bibinfo {pages} {184115} (\bibinfo {year}
  {2013})%
  \bibAnnoteFile{NoStop}{Bartok2013repr}%
\bibitem{Rupp2012}%
  \BibitemOpen
  \bibfield{author}{%
  \bibinfo {author} {\bibfnamefont{M.}~\bibnamefont{Rupp}}, \bibinfo {author}
  {\bibfnamefont{A.}~\bibnamefont{Tkatchenko}}, \bibinfo {author}
  {\bibfnamefont{K.-R.}\ \bibnamefont{M\"uller}},\ and\ \bibinfo {author}
  {\bibfnamefont{O.~A.}\ \bibnamefont{von Lilienfeld}},\ }%
  \bibfield{journal}{%
  \bibinfo {journal} {Phys. Rev. Lett.}\ }%
  \textbf{\bibinfo {volume} {108}},\ \bibinfo {pages} {058301} (\bibinfo {year}
  {2012})%
  \bibAnnoteFile{NoStop}{Rupp2012}%
\bibitem{behler2007}%
  \BibitemOpen
  \bibfield{author}{%
  \bibinfo {author} {\bibfnamefont{J.}~\bibnamefont{Behler}}\ and\ \bibinfo
  {author} {\bibfnamefont{M.}~\bibnamefont{Parrinello}},\ }%
  \bibfield{journal}{%
  \bibinfo {journal} {Phys. Rev. Lett.}\ }%
  \textbf{\bibinfo {volume} {98}},\ \bibinfo {pages} {146401} (\bibinfo {year}
  {2007})%
  \bibAnnoteFile{NoStop}{behler2007}%
\bibitem{behler2011atom}%
  \BibitemOpen
  \bibfield{author}{%
  \bibinfo {author} {\bibfnamefont{J.}~\bibnamefont{Behler}},\ }%
  \bibfield{journal}{%
  \bibinfo {journal} {J.~ Chem. Phys.}\ }%
  \textbf{\bibinfo {volume} {134}},\ \bibinfo {pages} {074106} (\bibinfo {year}
  {2011})%
  \bibAnnoteFile{NoStop}{behler2011atom}%
\bibitem{SNR83}%
  \BibitemOpen
  \bibfield{author}{%
  \bibinfo {author} {\bibfnamefont{P.}~\bibnamefont{Steinhardt}}, \bibinfo
  {author} {\bibfnamefont{D.}~\bibnamefont{Nelson}},\ and\ \bibinfo {author}
  {\bibfnamefont{M.}~\bibnamefont{Ronchetti}},\ }%
  \bibfield{journal}{%
  \bibinfo {journal} {Phys. Rev. B}\ }%
  \textbf{\bibinfo {volume} {28}},\ \bibinfo {pages} {784} (\bibinfo {year}
  {1983})%
  \bibAnnoteFile{NoStop}{SNR83}%
\bibitem{bartokthesis}%
  \BibitemOpen
  \bibfield{author}{%
  \bibinfo {author} {\bibfnamefont{A.~P.}\ \bibnamefont{Bart\'ok}},\ }%
  \emph{\bibinfo {title} {Gaussian Approximation Potential: an interatomic
  potential derived from first principles Quantum Mechanics}},\ Ph.D. thesis,\
  \bibinfo {school} {Cambridge} (\bibinfo {year} {2009})%
  \bibAnnoteFile{NoStop}{bartokthesis}%
\bibitem{flower1998}%
  \BibitemOpen
  \bibfield{author}{%
  \bibinfo {author} {\bibfnamefont{D.~R.}\ \bibnamefont{Flower}},\ }%
  \bibfield{journal}{%
  \bibinfo {journal} {J.~ Chem. Inf. Comp. Sci.}\ }%
  \textbf{\bibinfo {volume} {38}},\ \bibinfo {pages} {379} (\bibinfo {year}
  {1998})%
  \bibAnnoteFile{NoStop}{flower1998}%
\bibitem{Kabsch1976}%
  \BibitemOpen
  \bibfield{author}{%
  \bibinfo {author} {\bibfnamefont{W.}~\bibnamefont{Kabsch}},\ }%
  \bibfield{journal}{%
  \bibinfo {journal} {Acta Cryst A}\ }%
  \textbf{\bibinfo {volume} {32}},\ \bibinfo {pages} {922} (\bibinfo {year}
  {1976})%
  \bibAnnoteFile{NoStop}{Kabsch1976}%
\bibitem{coutsias2004}%
  \BibitemOpen
  \bibfield{author}{%
  \bibinfo {author} {\bibfnamefont{E.~A.}\ \bibnamefont{Coutsias}}, \bibinfo
  {author} {\bibfnamefont{C.}~\bibnamefont{Seok}},\ and\ \bibinfo {author}
  {\bibfnamefont{K.~A.}\ \bibnamefont{Dill}},\ }%
  \bibfield{journal}{%
  \bibinfo {journal} {J.~ Comput. Chem.}\ }%
  \textbf{\bibinfo {volume} {25}},\ \bibinfo {pages} {1849} (\bibinfo {year}
  {2004})%
  \bibAnnoteFile{NoStop}{coutsias2004}%
\bibitem{Theobald2005}%
  \BibitemOpen
  \bibfield{author}{%
  \bibinfo {author} {\bibfnamefont{D.~L.}\ \bibnamefont{Theobald}},\ }%
  \bibfield{journal}{%
  \bibinfo {journal} {Acta Cryst. A}\ }%
  \textbf{\bibinfo {volume} {61}},\ \bibinfo {pages} {478} (\bibinfo {year}
  {2005})%
  \bibAnnoteFile{NoStop}{Theobald2005}%
\bibitem{rose1995elementary}%
  \BibitemOpen
  \bibfield{author}{%
  \bibinfo {author} {\bibfnamefont{M.~E.}\ \bibnamefont{Rose}},\ }%
  \emph{\bibinfo {title} {Elementary theory of angular momentum}}\ (\bibinfo
  {publisher} {Courier Corporation},\ \bibinfo {year} {1995})%
  \bibAnnoteFile{NoStop}{rose1995elementary}%
\bibitem{varshalovich1988quantum}%
  \BibitemOpen
  \bibfield{author}{%
  \bibinfo {author} {\bibfnamefont{D.~A.}\ \bibnamefont{Varshalovich}},
  \bibinfo {author} {\bibfnamefont{A.}~\bibnamefont{Moskalev}},\ and\ \bibinfo
  {author} {\bibfnamefont{V.}~\bibnamefont{Khersonskii}},\ }%
  \emph{\bibinfo {title} {Quantum theory of angular momentum}}\ (\bibinfo
  {publisher} {World Scientific},\ \bibinfo {year} {1988})%
  \bibAnnoteFile{NoStop}{varshalovich1988quantum}%
\bibitem{SA1987}%
  \BibitemOpen
  \bibfield{author}{%
  \bibinfo {author} {\bibfnamefont{P.~J.}\ \bibnamefont{Van~Laarhoven}}\ and\
  \bibinfo {author} {\bibfnamefont{E.~H.}\ \bibnamefont{Aarts}},\ }%
  \emph{\bibinfo {title} {Simulated Annealing: Theory and Applications}},\
  \bibinfo {series} {Mathematics and its Applications}, Vol.~\bibinfo {volume}
  {37}\ (\bibinfo {publisher} {Springer Science \& Business Media},\ \bibinfo
  {year} {1987})%
  \bibAnnoteFile{NoStop}{SA1987}%
\bibitem{SA2005}%
  \BibitemOpen
  \bibfield{author}{%
  \bibinfo {author} {\bibfnamefont{J.~C.}\ \bibnamefont{Spall}},\ }%
  \emph{\bibinfo {title} {Introduction to Stochastic Search and Optimization:
  Estimation, Simulation, and Control}},\ \bibinfo {series} {Series in Discrete
  Mathematics and Optimization}, Vol.~\bibinfo {volume} {65}\ (\bibinfo
  {publisher} {John Wiley \& Sons},\ \bibinfo {year} {2005})%
  \bibAnnoteFile{NoStop}{SA2005}%
\bibitem{silverman1986density}%
  \BibitemOpen
  \bibfield{author}{%
  \bibinfo {author} {\bibfnamefont{B.~W.}\ \bibnamefont{Silverman}},\ }%
  \emph{\bibinfo {title} {Density Estimation for Statistics and Data
  Analysis}},\ Vol.~\bibinfo {volume} {26}\ (\bibinfo {publisher} {CRC press},\
  \bibinfo {year} {1986})%
  \bibAnnoteFile{NoStop}{silverman1986density}%
\bibitem{hardle2004nonparametric}%
  \BibitemOpen
  \bibfield{author}{%
  \bibinfo {author} {\bibfnamefont{W.}~\bibnamefont{H{\"a}rdle}},\ }%
  \emph{\bibinfo {title} {Nonparametric and Semiparametric Models}}\ (\bibinfo
  {publisher} {Springer Science \& Business Media},\ \bibinfo {year} {2004})%
  \bibAnnoteFile{NoStop}{hardle2004nonparametric}%
\bibitem{MRRTT53}%
  \BibitemOpen
  \bibfield{author}{%
  \bibinfo {author} {\bibfnamefont{N.}~\bibnamefont{Metropolis}}, \bibinfo
  {author} {\bibfnamefont{A.~W.}\ \bibnamefont{Rosenbluth}}, \bibinfo {author}
  {\bibfnamefont{M.~N.}\ \bibnamefont{Rosenbluth}}, \bibinfo {author}
  {\bibfnamefont{A.~H.}\ \bibnamefont{Teller}},\ and\ \bibinfo {author}
  {\bibfnamefont{E.}~\bibnamefont{Teller}},\ }%
  \bibfield{journal}{%
  \bibinfo {journal} {J. Chem. Phys.}\ }%
  \textbf{\bibinfo {volume} {21}},\ \bibinfo {pages} {1087} (\bibinfo {year}
  {1953})%
  \bibAnnoteFile{NoStop}{MRRTT53}%
\end{thebibliography}
\end{document}